\documentclass{article}

\usepackage{arxiv}
\usepackage[T1]{fontenc}
\newcommand{\authornamefont}{\normalfont\rmfamily\mdseries\fontsize{11}{13}\selectfont}
\newcommand{\affiliationfont}{\normalfont\rmfamily\mdseries\fontsize{9}{11}\selectfont}

\usepackage[square,numbers,sort&compress]{natbib}
\usepackage{amsmath,amssymb,amsfonts,bm}
\DeclareMathSizes{11}{10}{7}{5}
\usepackage{array}
\usepackage{booktabs}
\usepackage{arydshln}
\ADLinactivate
\usepackage{float}
\usepackage[bottom]{footmisc}
\usepackage{graphicx}
\usepackage{caption}
\usepackage{makecell}
\usepackage{microtype}
\usepackage{multirow}
\usepackage{needspace}
\usepackage{placeins}
\usepackage{threeparttable}
\usepackage{xcolor}
\usepackage{url}
\usepackage{hyperref}
\usepackage[nameinlink,noabbrev]{cleveref}

\graphicspath{{figures/}}
\hypersetup{hidelinks}
\newcommand{\reportlogo}{%
\includegraphics[height=26pt]{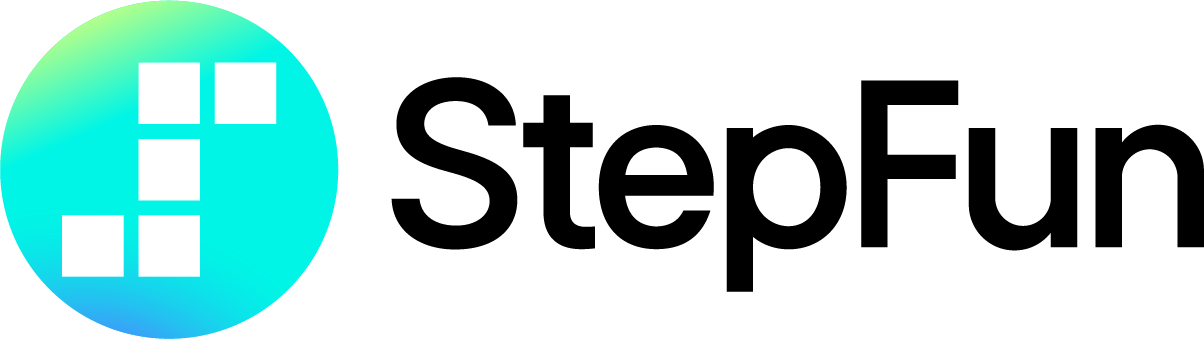}\hspace{18pt}%
\includegraphics[height=26pt]{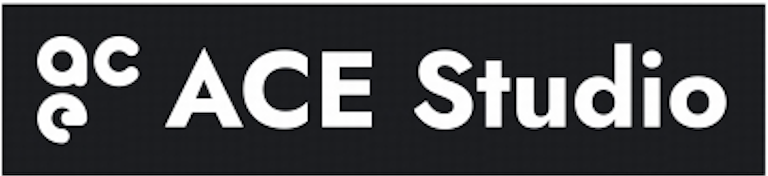}%
}
\title{StepAudio 3 Music Technical Report}

\author{%
    {\authornamefont Chengli Feng\textsuperscript{1}, Zhiyue Wu\textsuperscript{1}, Jiahao Song\textsuperscript{1}, Zheqi Dai\textsuperscript{1,3}, Boyang Wang\textsuperscript{1,4}, Ruibin Yuan\textsuperscript{2}, Junming Gong\textsuperscript{2},}\\[2pt]
    {\authornamefont Wenxiao Zhao\textsuperscript{2}, Jing Guo\textsuperscript{2}, Gang Yu\textsuperscript{1}, Xiangyu Zhang\textsuperscript{1}, Xuerui Yang\textsuperscript{1,}\thanks{Corresponding authors: \href{mailto:yangxuerui@stepfun.com}{yangxuerui@stepfun.com}, \href{mailto:yanchao@stepfun.com}{yanchao@stepfun.com}.}, Chao Yan\textsuperscript{1,}\footnotemark[1]}\\[6pt]
    {\affiliationfont \textsuperscript{1}StepFun\quad\textsuperscript{2}ACE\quad\textsuperscript{3}The Chinese University of Hong Kong\quad\textsuperscript{4}University of California San Diego}%
}

\renewcommand{\headeright}{}

\renewcommand{\shorttitle}{}

\begin{document}

\raggedbottom

\maketitle

\vspace{10pt}
\begin{abstract}
We introduce StepAudio 3 Music, a large-scale, long-form music generation model that supports explicit musical planning and open-domain text-controlled generation. The StepAudio Music Tokenizer represents audio as a \mbox{50-Hz} stream from a 65536-entry single codebook, using semantically informed self-supervised and multi-task training to preserve musical structure and reconstruction-relevant information. A flow-matching diffusion Transformer (DiT) predicts continuous StepAudio VAE latents, which our VAE decoder converts into 48-kHz audio. This discrete--continuous design is guided by comparisons of single-codebook VQ, Semantic and Acoustic RVQ, and different DiT configurations. For explicit planning, a Mixture-of-Experts autoregressive model uses ABC notation to produce an intermediate arrangement plan (ABC-CoT) before predicting music tokens, making harmony, rhythm, and melodic structure part of the generation context. A progressive training curriculum and supervised fine-tuning support song and instrumental generation, accompaniment generation from dry vocals, and cover-song synthesis for up to 5 minutes and 30 seconds. With reinforcement learning via direct preference optimization (DPO), the final model achieves the highest AudioBox Content Enjoyment, Content Usefulness, and Production Quality scores and the highest MuQ-MuLan similarity among the evaluated systems, with competitive SongBench results. On the preliminary Artificial Analysis Music Arena Vocals leaderboard, it obtains a Quality Elo of 1105, behind only Suno V5.5 and Mureka and ahead of Suno V5, MiniMax models, and other systems. Audio demonstrations are available at \mbox{\url{https://stepaudiollm.github.io/step-audio-3-music}}.
\end{abstract}

\begingroup
\setlength{\intextsep}{16pt}
\begin{figure}[H]
    \centering
    \includegraphics[width=0.74\textwidth]{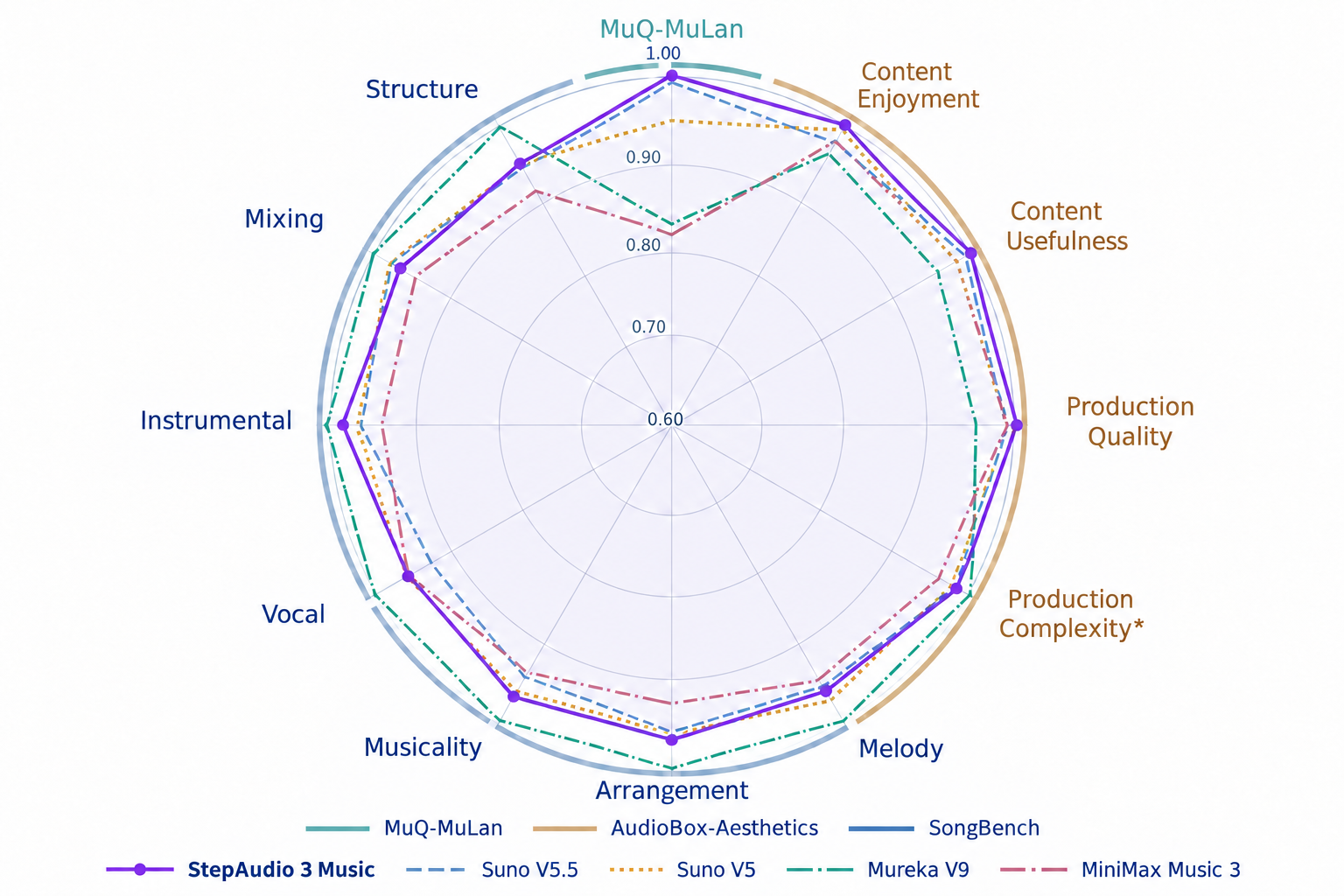}
    \caption{\textbf{Relative performance across MuQ-MuLan, AudioBox-Aesthetics, and SongBench.}
    Outer arcs identify the three benchmark families. Scores are divided by
    the best observed score per metric (1.0), not a theoretical maximum;
    the radial axis is truncated to 0.60--1.00.
    Purple denotes StepAudio 3 Music. Compare systems within each metric;
    polygon area is not an aggregate score. Curves provide a visual approximation;
    exact scores are in \cref{tab:objective-main}.}
    \label{fig:objective-radar}
\end{figure}
\endgroup
\clearpage

\section{Introduction}
\label{sec:introduction}

Recent years have seen substantial progress in the audio quality, duration, and task coverage of music generation models. Diffusion and flow-matching methods model complex acoustic distributions in continuous audio latent spaces, while diffusion Transformers (DiTs) provide an architecture for modeling long temporal contexts. Stable Audio's long-form latent diffusion model extended this approach to music with coherent full-length structure; DiffRhythm explored joint generation of vocals and accompaniment for complete songs; and ACE-Step further improved the efficiency of diffusion-based music synthesis \citep{evans2024longform,ning2025diffrhythm,gong2025acestep}. These advances have established a strong acoustic foundation for automatic music creation, bringing increasing attention to the organization of melody, arrangement, and long-range structure.

As generation moves from excerpts to complete songs, models must coordinate decisions at different scales: long-range musical choices about melodic development, section transitions, and the relationship between lyrics and music, alongside detailed acoustic realization of timbre, vocal texture, and transients. Hierarchical architectures combining language models with diffusion or flow-matching renderers provide an important route to this coordination. Seed-Music integrates autoregressive modeling and diffusion within a unified framework for conditioned song generation and editing \citep{bai2024seedmusic}. InspireMusic uses an autoregressive Transformer to predict single-codebook audio tokens, followed by a flow-matching model that supplies high-sampling-rate acoustic detail \citep{zhang2025inspiremusic}. Qwen-Music and ACE-Step 1.5 further explore the collaboration between language models and DiTs, assigning music sequence organization and high-fidelity rendering to different components \citep{xu2026qwenmusic,gong2026acestep15}. A recent full-song framework combines an eight-codebook RVQ tokenizer with hierarchical autoregressive token modeling and FullDiT, a flow-matching renderer operating in continuous VAE latent space \citep{dai2026fullsong}. This division of labor supports long-form generation and creates an opportunity to introduce musical planning before acoustic synthesis.

Music chain-of-thought (CoT) approaches make part of this planning explicit. MusiCoT, introduced by Kunlun, constructs an intermediate sequence of musical thoughts from quantized CLAP representations before generating audio tokens, allowing analysis of properties such as instrumental arrangement \citep{lam2025musiccot}. Qwen-Music introduces Melody-CoT, which uses melody tokens to plan vocal melodic contours before full-song generation and provides melody conditioning for cover synthesis \citep{xu2026qwenmusic}. ACE-Step 1.5 uses a language model to generate musical metadata, lyrics, and captions that guide subsequent synthesis \citep{gong2026acestep15}. Together, these approaches illustrate an expanding role for intermediate representations: beyond connecting model components, they can explicitly guide musical content and structure.

For controllable music creation, such a plan should also be understandable and actionable for the creator. Global text prompts describe genre, mood, and instrumentation, whereas specific creative decisions often concern how a phrase develops, where a chord changes, or how a chorus melody relates to a verse. Prior work has explored these musical elements as direct control conditions. JASCO combines global text descriptions with local conditions such as chords, melodies, and drum tracks to provide temporally grounded control \citep{tal2024jasco}; Seed-Music supports score conditioning and subsequent editing of lyrics and vocal melodies \citep{bai2024seedmusic}. InstructME supports instruction-guided music editing and remixing with latent diffusion, using multi-scale source features and chord conditioning to promote consistency and harmony \citep{han2023instructme}. These capabilities motivate a further question: how can a model's own intermediate plan expose clear musical semantics and temporal structure, so that creators can inspect and revise its arrangement decisions before audio is generated?

Symbolic music offers a natural language for this interface. ChatMusician uses text-compatible ABC notation for music understanding and generation, demonstrating that language models can directly work with scores and musical conditions such as chords, melodies, and form \citep{yuan2024chatmusician}. SongComposer jointly models lyrics and melodies through explicit representations of lyrics, pitch, duration, and rests \citep{ding2024songcomposer}. These studies establish a basis for expressing musical content as structured sequences that language models can process and creators can read. Such representations can complement open-domain text conditioning with an explicit planning interface, provided that the generation model can connect musical structure to its acoustic realization.

We introduce StepAudio 3 Music, a large-scale, long-form music generation model that supports explicit musical planning and open-domain text-controlled generation. The system combines a generation-oriented music tokenizer, a Mixture-of-Experts (MoE) autoregressive model, and a flow-matching DiT renderer. Lyrics, text prompts, and task-specific references provide creative conditions, while explicit planning supplies a structured musical context when enabled. As illustrated in \cref{fig:system-overview}, these components separate music representation, sequence modeling, and acoustic realization.

A reliable audio representation is fundamental to long-form generation. We conduct a controlled comparison of single-codebook vector quantization (VQ), Semantic residual vector quantization (RVQ), and Acoustic RVQ at a common rate of 25~Hz. Multi-codebook RVQ provides stronger acoustic reconstruction, whereas single-codebook VQ offers a more predictable autoregressive target and supports more stable musical development. This finding highlights a generation-oriented trade-off: a representation must preserve acoustic information while enabling the language model to maintain musical consistency over long sequences. The representation that reconstructs best is therefore not necessarily the one best suited to autoregressive generation.

Based on this finding, the StepAudio 3 Music Tokenizer adopts a 50-Hz single-codebook representation with 65536 entries, providing denser temporal supervision while retaining a single token stream. We train the tokenizer with semantically informed self-supervised and multi-task objectives to preserve both musical information and reconstruction-relevant detail. A flow-matching DiT then predicts continuous StepAudio VAE latents, which our VAE decoder converts into 48-kHz waveform audio. This division of labor lets the autoregressive model focus on musical sequence organization while the continuous renderer supplies acoustic detail.

For explicit musical planning, we use ABC notation to express an intermediate arrangement plan, referred to as ABC-CoT. In this mode, the MoE model first organizes chords, tempo, meter, key, bar structure, and melody into a readable plan, then predicts music tokens conditioned on it. ABC-CoT complements global text instructions with a temporally structured context and exposes arrangement decisions for inspection and revision before synthesis. Training on music-to-ABC and ABC-to-music tasks connects this symbolic representation to audio. The goal is to translate musical plans into coherent performances; adherence to individual notes, chords, and bars is distinct from the caption alignment measured in our current evaluation.

The model follows a progressive, multi-task training curriculum. Large-scale pre-training establishes general music generation, recognition, and understanding capabilities; multi-task mid-training introduces explicit planning and task-specific reference conditioning; and high-quality annealing focuses the training distribution on core creation tasks. Supervised fine-tuning strengthens task execution, and Direct Preference Optimization (DPO) aligns outputs with expert judgments of condition adherence and musical quality. The resulting system supports song generation, instrumental generation, accompaniment generation from dry vocals, and cover-song synthesis for durations of up to 5 minutes and 30 seconds, with text, notation, and reference conditions supporting complementary forms of creative control.

\Needspace{6\baselineskip}
Our main contributions are as follows:
\begin{itemize}
    \item \textbf{A generation-oriented music representation and renderer.} Controlled 25-Hz comparisons of single-codebook VQ and Semantic and Acoustic RVQ motivate a 50-Hz, 65536-entry single-codebook StepAudio Music Tokenizer. Semantically informed training preserves musical structure and acoustic information, while a flow-matching DiT and our StepAudio VAE provide 48-kHz rendering.
    \Needspace{5\baselineskip}
    \item \textbf{Text-controlled generation with explicit musical planning.} A MoE autoregressive model supports open-domain text conditions and task-specific references. For planning, ABC-CoT expresses harmony, rhythm, melody, and form as a readable intermediate arrangement before music-token prediction. Music-to-ABC and ABC-to-music training connect the symbolic plan to its acoustic realization.

\end{itemize}

\section{Model Architecture}
\label{sec:architecture}

\subsection{System Overview}

\begin{figure}[!htbp]
    \centering
    \includegraphics[width=0.96\textwidth]{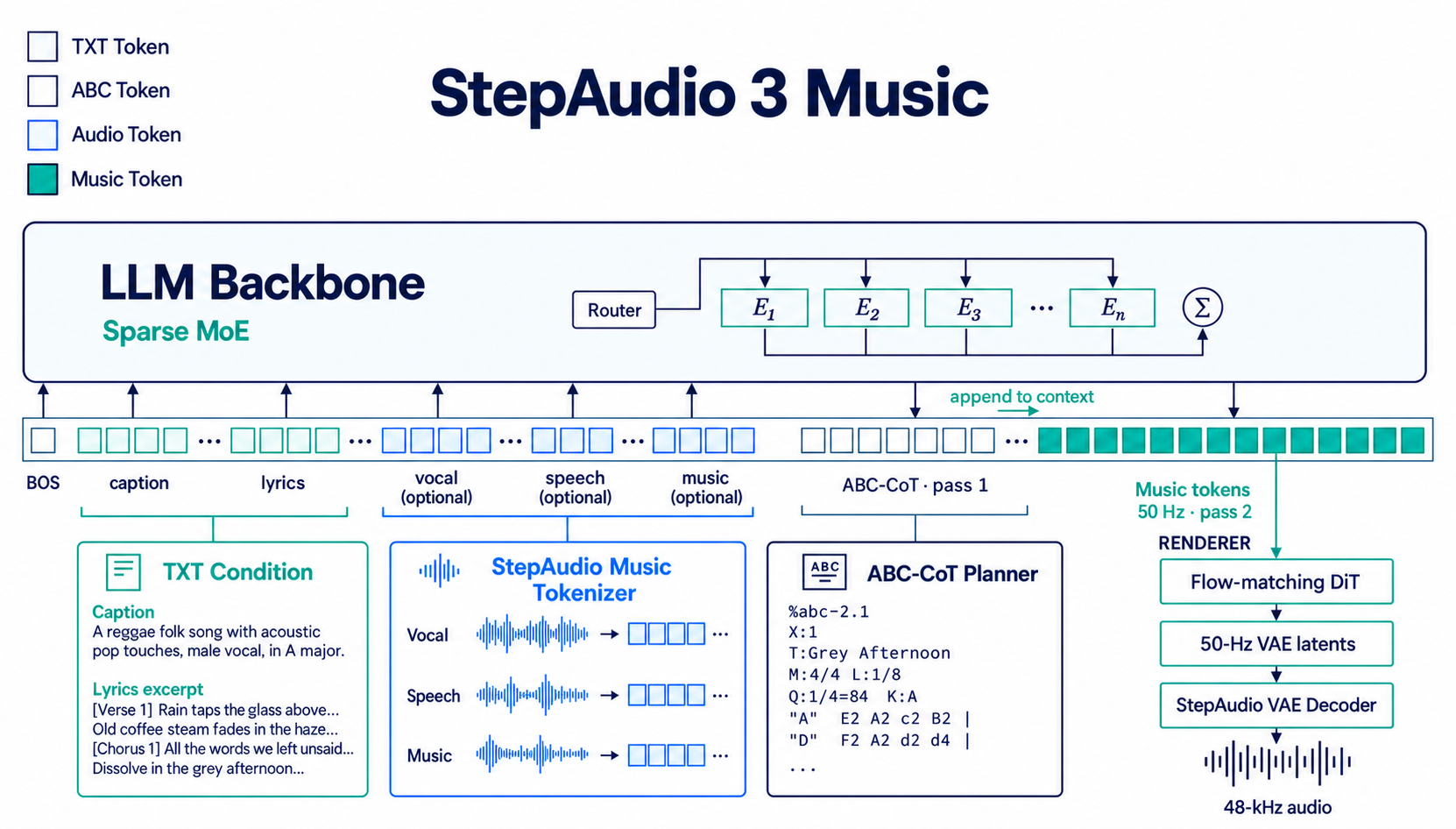}
    \caption{\textbf{Overview of StepAudio 3 Music.}
    Lyrics, a text prompt, and optional task-specific references are serialized for a trainable Mixture-of-Experts autoregressive model. In ABC-CoT mode, the model first produces an explicit arrangement plan and then predicts a 50-Hz sequence from a 65536-entry single codebook. A separately trained renderer generates 50-Hz StepAudio VAE latents with a flow-matching DiT and decodes them into 48-kHz waveform audio. The renderer is held fixed during autoregressive-model training.}
    \label{fig:system-overview}
\end{figure}

StepAudio 3 Music supports open-domain text-controlled generation and explicit musical planning through a shared autoregressive backbone. As shown in \cref{fig:system-overview}, a Mixture-of-Experts decoder conditions on lyrics, a text prompt, and optional task-specific references to predict music tokens. The StepAudio Music Codec provides the interface between this discrete sequence and high-fidelity audio, separating long-range musical organization from acoustic rendering.

The StepAudio Music Tokenizer represents audio with one discrete token per frame at 50~Hz from a 65536-entry single codebook. A flow-matching DiT detokenizer maps the token sequence to continuous StepAudio VAE latents, after which our frozen VAE decoder reconstructs 48-kHz waveform audio. The codec is trained independently and held fixed while the autoregressive music model is optimized, allowing the two stages to specialize in sequence prediction and acoustic realization, respectively.

When explicit planning is enabled, the autoregressive model first produces an ABC-CoT arrangement plan and incorporates it into the context for music-token generation. This mode complements text conditioning with a temporally structured musical representation, as detailed in \cref{sec:abc-control}.

\Needspace{6\baselineskip}
\subsection{Explicit Musical Planning with ABC Notation}
\label{sec:abc-control}

For explicit musical planning, the LLM uses ABC notation to represent an intermediate arrangement. The ABC-CoT plan combines global attributes, including tempo, meter, and key, with chords, bar structure, and a melodic sequence. These fields specify musical content at distinct temporal scales: global attributes establish the musical setting, while the chord and note sequence describes how it develops over successive bars.

Let $c$ denote the lyrics, text prompt, and optional reference conditions, $a$ the ABC-CoT plan, and $m_{1:T}$ the sequence of music tokens. The MoE model uses the following two-pass factorization:
\begin{equation}
    p_{\theta}(a,m_{1:T}\mid c)
    = p_{\theta}(a\mid c)
      \prod_{t=1}^{T} p_{\theta}(m_t\mid c,a,m_{<t}).
    \label{eq:abc-generation}
\end{equation}
In the first pass, the model produces the arrangement plan. The completed plan is appended to the context, and the second pass predicts the 50-Hz music-token sequence. Both passes use the same autoregressive backbone. The renderer then realizes the predicted sequence acoustically.

This factorization makes ABC notation a conditioning interface for melody, harmony, rhythm, and form. In text-and-lyrics generation, the model produces the symbolic plan from the requested conditions; in ABC-to-music generation, the ABC sequence itself supplies the musical condition. Training on both music-to-ABC and ABC-to-music tasks connects symbolic descriptions with their acoustic realizations, as detailed in \cref{sec:pretraining}. ABC-CoT therefore places explicit musical decisions in the generation context before acoustic synthesis. The degree to which the generated audio follows those decisions remains an empirical question, distinguished from caption alignment in \cref{sec:evaluation}.

\subsection{Music Codec}
\label{sec:music-codec}

Our codec study follows a bottleneck-first diagnosis guided by end-to-end generation quality rather than reconstruction fidelity alone. We first hold the discrete representation fixed and ask whether a stronger continuous renderer---through a higher-capacity VAE latent space or a larger DiT---raises final quality. The detokenizer analysis below separates direct VAE reconstruction from token-conditioned rendering and complements the reported objective metrics with internal perceptual comparisons. Across these analyses, gains from a stronger VAE representation or a larger DiT do not translate consistently to the complete token-conditioned pathway. This behavior identifies the discrete representation as an effective ceiling under the tested renderer configurations and motivates the tokenizer and bottleneck study that follows.

The StepAudio Music Codec provides the interface between continuous audio and the autoregressive model. Its discrete representation must satisfy two competing requirements. First, it must preserve sufficient information for the detokenizer to synthesize high-fidelity music. Second, it must form a stable and predictable sequence from which the language model can learn melody, rhythm, harmony, and vocal content. A codec optimized exclusively for reconstruction may devote much of its capacity to timbre, transients, and local spectral variation. Semantically similar passages can have many acoustically valid realizations, so preserving every local detail can increase the conditional uncertainty of token prediction. Conversely, an excessively compressed semantic representation can simplify language modeling while limiting the quality of acoustic reconstruction. Our codec is designed to balance these objectives.

The codec comprises a music tokenizer and a DiT-based detokenizer. The tokenizer converts 24-kHz audio into a 50-Hz stream of discrete tokens. Conditioned on these tokens, the detokenizer generates continuous VAE latents, which are decoded into 48-kHz waveform audio. We analyze this chain from downstream to upstream: first the VAE latent representation and DiT capacity, then the discrete bottleneck and the selected tokenizer architecture. Their training procedures are collected in \cref{sec:codec-training}.

\subsection{Flow-Matching Music Detokenizer}
\label{sec:music-detokenizer}

The StepAudio Music Detokenizer generates continuous VAE latents instead of predicting waveform samples directly. Conditioned on the 50-Hz music tokens, a DiT produces the 50-Hz, 64-channel latent representation of our StepAudio VAE; the frozen StepAudio VAE decoder then converts the generated latents into 48-kHz waveform audio. The token sequence and VAE latents share the same frame rate and are aligned frame by frame before the discrete condition is concatenated with the DiT input.

Long-form generation is performed in 30-second chunks. The first chunk begins with an all-zero two-second VAE-latent context; each subsequent chunk is conditioned on the final two seconds of the latent sequence generated for the preceding chunk. This recurrent context transfers local acoustic state across chunk boundaries and supports continuous long-form rendering. The corresponding training examples and conditioning procedure are described in \cref{sec:dit-training}.

For each target segment, the token sequence is the only conditioning signal that spans its complete duration; the two-second VAE-latent context carries only local continuity from the preceding audio. Content absent from both the token stream and this short context therefore cannot be recovered reliably by increasing renderer capacity alone. A stronger conditional generator can learn a more expressive token-to-latent mapping and can improve the plausibility of synthesized details, but it cannot reconstruct source-specific information that its conditioning does not preserve. We hold the tokenizer fixed, use its ground-truth tokens to avoid confounding the analysis with autoregressive prediction errors, and study two possible downstream bottlenecks: the VAE latent representation and the capacity of the DiT. The quantitative results are collected in the detokenizer ablation in \cref{subsec:detokenizer-ablation} (\cref{tab:music-detokenizer-ablations}).

The native 50-Hz StepAudio VAE achieves better direct encode--decode reconstruction than the 25-Hz StepAudio VAE. We therefore use the 50-Hz latent representation for the token-conditioned renderer. At this fixed rate, scaling the DiT from 0.9B to 4B or 8B parameters does not produce a consistent gain: the 0.9B model gives the best MCD, MS-Mel-L1, MS-STFT-L1, SI-SNR, and UTMOS, while the 4B model gives the best SDR. These results indicate an effective information ceiling imposed by the tokenizer under the tested renderer configurations. We adopt the 0.9B DiT and shift the design focus upstream to the tokenizer.

\subsection{Discrete Bottleneck Design}
\label{sec:discrete-bottleneck}

\begin{figure}[!htbp]
    \centering
    \includegraphics[width=0.94\textwidth]{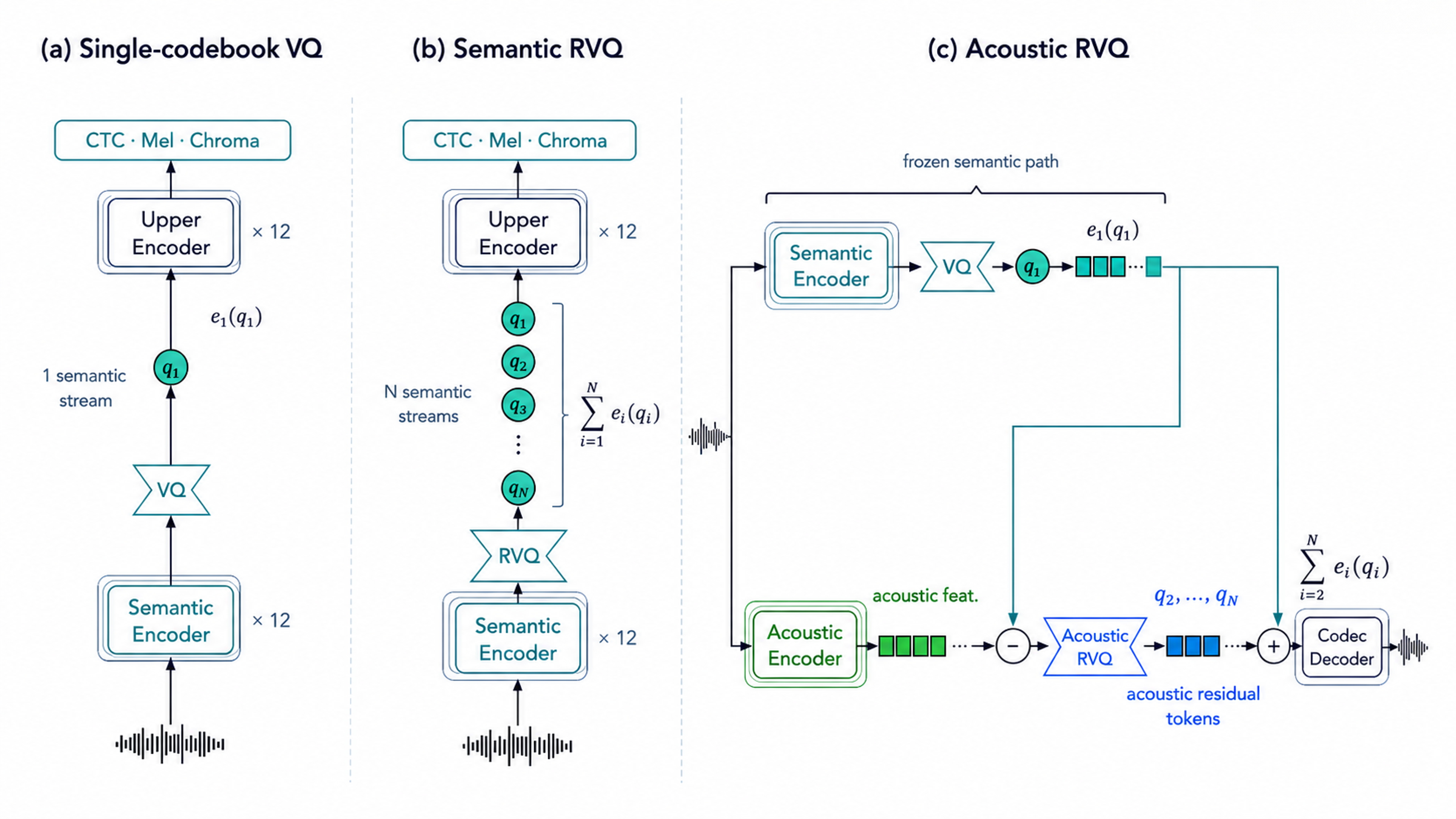}
    \caption{\textbf{Discrete bottleneck designs considered for the music tokenizer.}
    The controlled comparison uses a common 25-Hz frame rate for all three variants.
    (a) Single-codebook VQ produces one semantic token $q_1$ per frame.
    (b) Semantic RVQ quantizes residuals within the same semantically supervised latent space; the aggregate embedding $\sum_i e_i(q_i)$ is processed by the upper encoder under the shared multi-task objectives.
    (c) Acoustic RVQ freezes the semantic tokenizer and quantizes the residual between an acoustic feature and the aligned semantic feature.
    Teal elements denote semantic representations and tokens, green denotes acoustic features, and blue denotes acoustic residual tokens. A separate DiT detokenizer is trained for the final comparison after each tokenizer is frozen and is omitted from the diagram.}
    \label{fig:music-tokenizer-designs}
\end{figure}

Motivated by the effective ceiling observed in the detokenizer study, we compare the three candidate bottlenecks summarized in \cref{fig:music-tokenizer-designs}: single-codebook VQ, Semantic RVQ, and Acoustic RVQ. To isolate the effect of bottleneck structure, all three candidates operate at a common 25-Hz frame rate. They use the same bidirectional Conformer backbone and multi-task supervision described in \cref{sec:tokenizer-training}, keeping sequence length and training conditions comparable across variants. Each candidate tokenizer is frozen and paired with an independently trained DiT detokenizer. Reconstruction from ground-truth tokens measures the recoverable information in the discrete representation, while FastLM token-prediction accuracy and reconstruction from predicted tokens measure language-model predictability and end-to-end generation quality.

\paragraph{Single-codebook VQ.}
Single-codebook VQ produces one token $q_1$ for every 25-Hz frame in the controlled comparison. Its embedding $e_1(q_1)$ is passed through the upper Conformer layers and optimized by the CTC, Mel, and Chroma objectives. Because the language model predicts only one token stream, this representation has the simplest prediction target and the shortest error-propagation path.

\paragraph{Semantic RVQ.}
Semantic RVQ replaces the single VQ module at the same bottleneck. It sequentially quantizes the residual representation into $N$ token streams, $q_1,\ldots,q_N$, and aggregates their codebook-specific embeddings as
\begin{equation}
    \mathbf{z}_{\mathrm{RVQ}}
    = \sum_{i=1}^{N} e_i(q_i).
    \label{eq:semantic-rvq-sum}
\end{equation}
The aggregate is processed by the upper Conformer layers and trained with the same CTC, Mel, and Chroma objectives. Consequently, all codebooks are jointly optimized within the shared, semantically supervised representation space rather than being dedicated exclusively to low-level acoustic residuals.

\paragraph{Acoustic RVQ.}
Acoustic RVQ adapts the dual-stream principle introduced for speech coding in DualCodec \citep{li2025dualcodec}. We first train and freeze a single-codebook semantic tokenizer that produces $q_1$ and its embedding $e_1(q_1)$. A separate acoustic encoder extracts a continuous acoustic feature from the waveform. Additional RVQ codebooks encode the residual between this feature and the aligned semantic representation into $q_2,\ldots,q_N$. During decoding, the residual embeddings $\sum_{i=2}^{N}e_i(q_i)$ are added to the semantic embedding and passed to the codec decoder. The acoustic residual streams are optimized primarily for waveform reconstruction and are not directly constrained by the semantic objectives.

Multi-layer RVQ is commonly used to increase the capacity of high-fidelity music codecs. MiniMax Music~3, for example, uses eight codebooks: a first codebook with 16{,}384 entries for core musical structure, followed by seven 1{,}024-entry residual codebooks for progressively finer acoustic detail \citep{minimax2026music3}. This design illustrates the representational capacity of RVQ, but the number and size of its codebooks do not by themselves determine whether the tokens are easy for an autoregressive model to predict. The placement of semantic supervision relative to residual quantization is equally important.

At the common 25-Hz rate, both Semantic RVQ and Acoustic RVQ provide substantially better reconstruction fidelity than single-codebook VQ when decoded from ground-truth tokens. This advantage does not necessarily carry over to generation from predicted tokens. Acoustic RVQ is difficult for FastLM to predict: unlike the features used by Semantic RVQ, its acoustic residual features do not undergo pre-clustering. In subjective listening evaluations by our internal music experts, generations based on Acoustic RVQ are judged poor in both audio fidelity and musicality.

FastLM predicts Semantic RVQ more reliably, producing generated audio with good fidelity. However, the same internal music experts find that its musicality remains weak. We attribute this degradation to error accumulation across residual codebook streams: prediction errors are combined when embeddings are aggregated, and errors in earlier codebooks can affect subsequent predictions. In contrast, single-codebook VQ incurs only a modest reduction in generated audio fidelity relative to Semantic RVQ while maintaining good musicality, as assessed by the same experts. These comparisons motivate selecting single-codebook VQ for generation rather than choosing a bottleneck solely for its reconstruction quality.

After selecting the single-codebook design, we conduct a separate 25-Hz versus 50-Hz ablation using the same dataset. We observe no noticeable difference in the audio fidelity of the generated music, but the 50-Hz configuration improves musicality. We hypothesize that this benefit arises from the denser token-level supervision: for the same amount of music, a 50-Hz representation provides twice as many training tokens as a 25-Hz representation. It therefore increases the number of prediction targets available from the existing corpus, rather than adding new musical material. Given the scarcity of music training data, we adopt 50~Hz with a 65536-entry single codebook as the final model setting. Although this choice doubles the autoregressive sequence length, it retains the single-stream prediction structure favored by the 25-Hz bottleneck comparison. The resulting tokenizer architecture is described below, and its training objectives are detailed in \cref{sec:tokenizer-training}.

\subsection{StepAudio Music Tokenizer}
\label{sec:music-tokenizer}

The StepAudio Music Tokenizer uses a Mel frontend followed by a 24-layer bidirectional Conformer \citep{gulati2020conformer}. The frontend converts 24-kHz audio into a 100-Hz Mel representation and subsamples it to 50~Hz. The model contains approximately 0.6 billion parameters, with a hidden dimension of 1{,}024. A single-codebook VQ bottleneck sits between the lower and upper 12-layer Conformer stacks. The lower stack projects its output into a 32-dimensional space for quantization with 65536 entries; the quantized embeddings are projected back to the hidden dimension before the upper stack. The three-stage optimization of this architecture is described in \cref{sec:tokenizer-training}.

The resulting tokenizer emits one token per frame at 50~Hz. Because the codebook contains $65536=2^{16}$ entries, each token can be represented by 16 bits. The corresponding nominal bitrate is therefore
\begin{equation}
    50~\mathrm{frames/s} \times 16~\mathrm{bits/frame}
    = 800~\mathrm{bit/s},
    \label{eq:tokenizer-bitrate}
\end{equation}
excluding entropy coding and auxiliary side information.

\Needspace{7\baselineskip}
\section{Data Pipeline}
\label{sec:data-pipeline}

\begin{figure}[!htbp]
    \centering
    \includegraphics[width=0.94\textwidth]{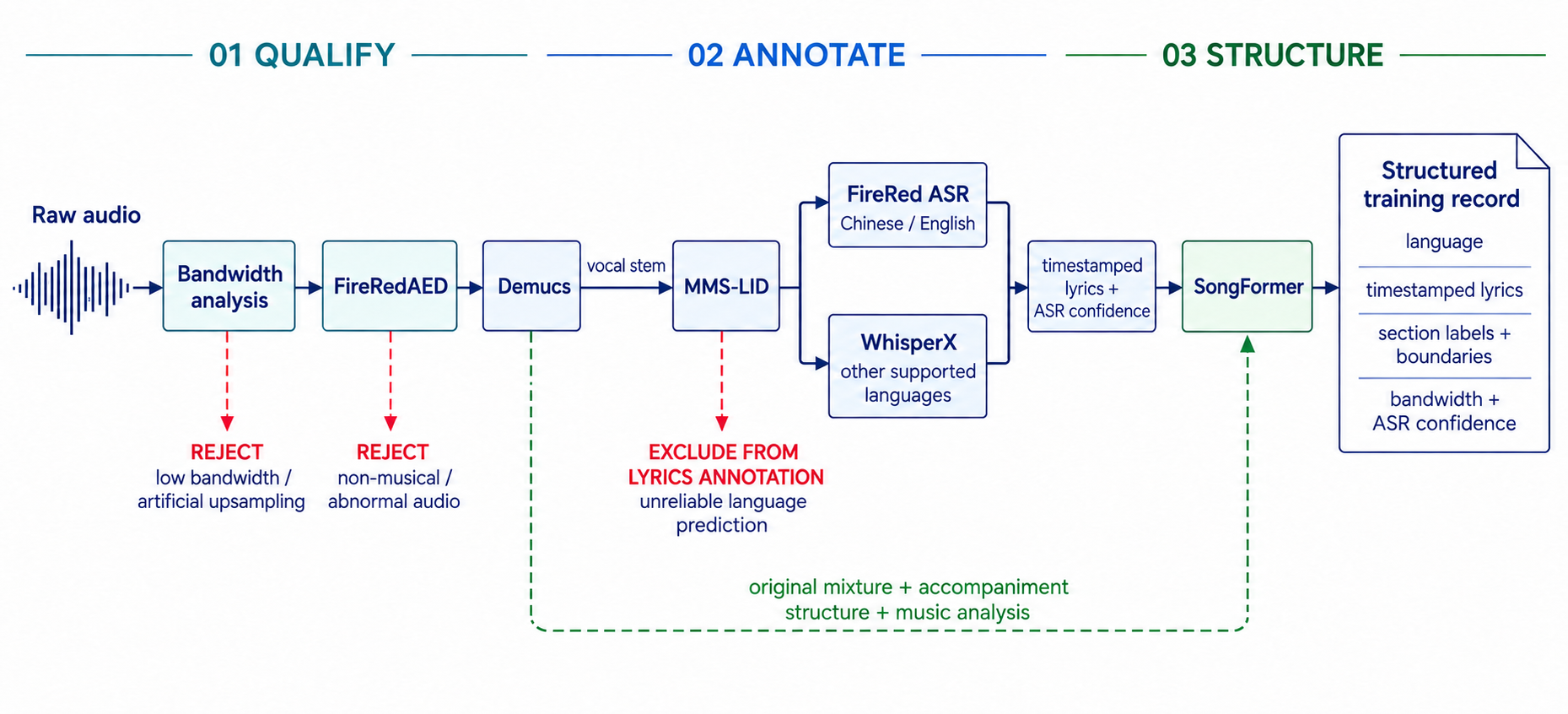}
    \caption{\textbf{Music data cleaning and structured annotation pipeline.}
    Raw audio is filtered for bandwidth and content quality, separated into vocal and accompaniment stems, routed to language-appropriate transcription systems, and aligned with predicted song sections to produce structured training records. Dashed branches indicate rejection or auxiliary data paths.}
    \label{fig:data-pipeline}
\end{figure}

Large-scale music training data must satisfy both acoustic and structural quality requirements. Music samples vary in bandwidth, content type, language, transcription quality, and structural metadata. We therefore construct an automated pipeline that progressively filters low-quality material and converts each retained song into a structured training record. As illustrated in \cref{fig:data-pipeline}, the pipeline performs frequency-band analysis, audio-event detection, source separation, language identification, multilingual lyrics transcription, and song-structure analysis.

\subsection{Acoustic Quality and Content Filtering}

We first analyze the spectrum of each audio file to estimate its effective bandwidth. This model-free stage identifies low-bandwidth recordings and likely artificial upsampling. The resulting bandwidth measurements provide a basic indicator of technical audio quality and are used to remove samples with severely degraded or insufficient spectral information.

We then perform multi-label AED using the audio event detection module of \texttt{FireRedVAD}, provided as part of \texttt{FireRedASR2S} \citep{xu2026fireredasr2s}. The module detects speech, singing, and music, producing event timestamps and posterior probabilities. These predictions support content filtering to exclude non-musical recordings and samples with insufficient musical content before source separation and lyrics transcription. This early filtering step reduces downstream computation and limits the propagation of irrelevant material through the annotation pipeline.

\subsection{Source Separation and Multilingual Lyrics Transcription}

For each retained sample, we use \texttt{Demucs} \citep{defossez2019music} to separate the original mixture into vocal and accompaniment tracks. The isolated vocal track is used for language identification and lyrics transcription, while the original mixture and accompaniment are retained for structural and music-related analysis. Performing recognition on separated vocals reduces interference from instrumentation and improves the reliability of both language and lyrics predictions.

We identify the primary vocal language using \texttt{MMS-LID} \citep{pratap2023scaling}. Samples with sufficiently reliable language predictions are routed to a language-appropriate automatic speech recognition system; samples with uncertain predictions are excluded from downstream lyrics annotation. Chinese- and English-language songs are transcribed with \texttt{FireRed ASR} \citep{xu2025fireredasr}, while other supported languages are processed with \texttt{WhisperX} \citep{bain2023whisperx}. This routing strategy exploits the complementary strengths of the two recognition systems and improves the robustness of multilingual transcription.

For every recognized segment, we retain the transcribed text, start and end timestamps, and ASR confidence. Segment timestamps later align lyrics with predicted song sections, while confidence scores provide a quality signal for filtering and task-specific data selection. Depending on the downstream objective, samples with low confidence or poor lyrics--audio alignment are either removed or assigned explicit quality labels. This policy limits annotation noise while preserving corpus scale and musical diversity.

\subsection{Song Structure and Training Records}

After obtaining timestamped lyrics, we apply \texttt{SongFormer} \citep{hao2025songformer} to analyze the global structure of each complete song. The model partitions the audio into sections such as introductions, verses, choruses, bridges, and outros. Each lyrics segment is assigned to the corresponding section according to its temporal overlap with the predicted boundaries.

The final record contains the detected language, timestamped lyrics, section labels and boundaries, frequency-band measurements, and ASR confidence scores. These structured annotations support subsequent filtering and provide supervision for lyrics-to-song generation, song-to-lyrics recognition, and long-form song-structure modeling. By retaining both annotations and their associated confidence signals, the pipeline enables different training stages to apply task-specific quality thresholds without discarding the underlying corpus prematurely.

\section{Training}
\label{sec:training}

Training proceeds from the audio representation to the language model. We first train the StepAudio Music Tokenizer and the DiT renderer, then hold the codec fixed while optimizing the MoE language model. LLM pretraining develops text-conditioned generation and explicit musical planning through a progressive multi-task curriculum connecting text, ABC notation, and music tokens. Post-training specializes the model for music creation and aligns its outputs with expert listening preferences.

\subsection{Tokenizer \& DiT Training}
\label{sec:codec-training}

The audio codec is trained independently of the LLM. The tokenizer learns a discrete representation through three successive stages, after which its parameters are frozen for token-conditioned DiT training. The final codec uses a 50-Hz, 65536-entry single-codebook representation; the 25-Hz variants in \cref{sec:discrete-bottleneck} serve only as controlled comparisons of bottleneck structure.

\subsubsection{Tokenizer Training}
\label{sec:tokenizer-training}

\begin{figure}[!htbp]
    \centering
    \includegraphics[width=0.96\textwidth]{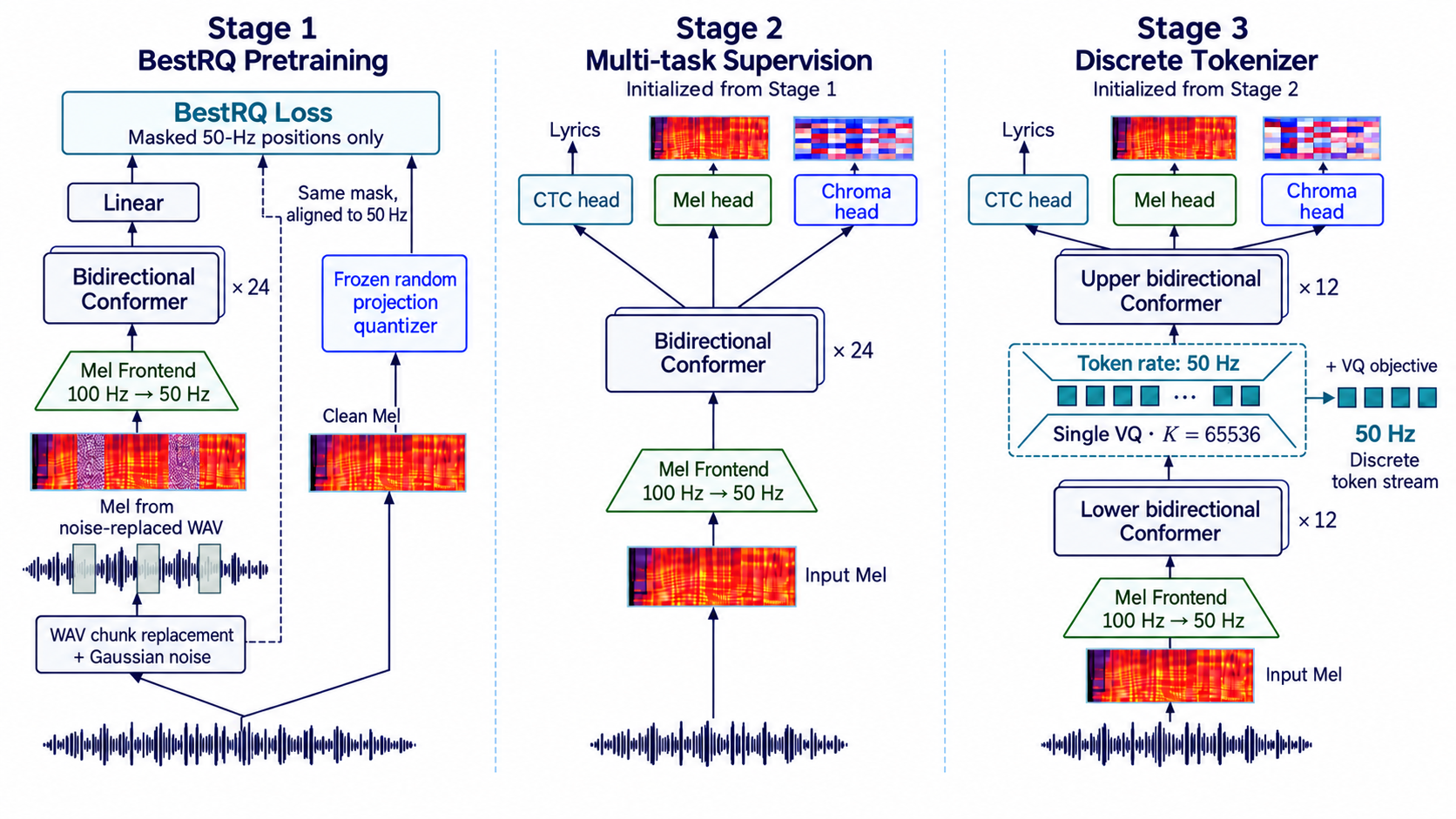}
    \caption{\textbf{Three-stage training pipeline for the StepAudio Music Tokenizer.}
    Stage~1 adapts BEST-RQ pretraining with bidirectional Conformer blocks. Contiguous waveform spans are replaced with Gaussian noise before Mel extraction, and the BEST-RQ loss is evaluated only at the corresponding positions in the aligned 50-Hz sequence. Stage~2 introduces full-song CTC, Mel, and Chroma supervision. Stage~3 inserts a single-codebook VQ bottleneck with 65536 entries between the lower and upper 12-layer Conformer stacks, producing one discrete music token per frame at 50~Hz. Each stage is initialized from the checkpoint produced by the preceding stage.}
    \label{fig:music-tokenizer-training}
\end{figure}

We progressively introduce musical supervision and discretization into the Conformer architecture described in \cref{sec:music-tokenizer}. \Cref{fig:music-tokenizer-training} summarizes the three stages and the checkpoint initialization between them.

\paragraph{Stage~1: Self-Supervised Pretraining.}

We adapt the BEST-RQ self-supervised objective \citep{chiu2022bestrq} for music pretraining. In our implementation, the input waveform is processed through an online branch and a clean target branch. In the online branch, randomly selected contiguous waveform spans are replaced with Gaussian noise before Mel extraction. The corresponding Mel regions therefore contain broadband stochastic energy rather than zeroed or directly masked Mel bins. These corrupted features are passed through the frontend and the 24-layer bidirectional Conformer. In the target branch, clean Mel features from the uncorrupted waveform are processed by a frozen random-projection quantizer to produce discrete targets. The model predicts the clean targets, and the BEST-RQ objective is evaluated only at 50-Hz positions aligned with the waveform-domain corruption mask.

\paragraph{Stage~2: Multi-Task Representation Learning.}

Inspired by the progressive separation of representation learning and discretization in Duo-Tok \citep{lin2025duotok}, Stage~2 is initialized from Stage~1 and introduces multi-task supervision over full-song inputs. A CTC head uses normalized lyrics to provide linguistic and vocal-content supervision. A Mel reconstruction objective encourages the representation to retain timbre, energy, transients, and local spectral structure, while a Chroma reconstruction objective promotes the preservation of pitch class, tonality, and harmonic organization. Together, these objectives adapt the self-supervised representation into a structured space suitable for music tokenization.

\paragraph{Stage~3: Discrete Tokenization.}

Stage~3 is initialized from Stage~2 and inserts a discrete bottleneck between the 12th and 13th Conformer blocks. The lower 12 layers encode the audio into a continuous music representation, which is projected into a 32-dimensional space and quantized using a single codebook with 65536 entries. The quantized embeddings are projected back to the Conformer hidden dimension, processed by the upper 12 layers, and trained with the same CTC, Mel, and Chroma heads. Abstractly, the joint training objective is
\begin{equation}
    \mathcal{L}_{\mathrm{tokenizer}}
    = \lambda_{\mathrm{CTC}}\mathcal{L}_{\mathrm{CTC}}
    + \lambda_{\mathrm{Mel}}\mathcal{L}_{\mathrm{Mel}}
    + \lambda_{\mathrm{Chroma}}\mathcal{L}_{\mathrm{Chroma}}
    + \lambda_{\mathrm{VQ}}\mathcal{L}_{\mathrm{VQ}},
    \label{eq:tokenizer-objective}
\end{equation}
where the four terms denote lyric recognition, Mel reconstruction, Chroma reconstruction, and vector-quantization objectives, respectively. This combination encourages each token to retain both musically structured content and reconstruction-relevant acoustic information.

\subsubsection{DiT Training}
\label{sec:dit-training}

Once the tokenizer is trained, we freeze it and train the flow-matching DiT on its ground-truth music tokens and the corresponding continuous latents from the StepAudio VAE. The discrete condition is aligned frame by frame with the 50-Hz VAE latents and concatenated with the DiT input. The StepAudio VAE decoder is held fixed. This stage learns acoustic rendering independently of autoregressive token-prediction errors.

Each example contains a two-second VAE-latent context followed by a 30-second target segment. A segment at the beginning of a song uses an all-zero context; other segments use ground-truth VAE latents from the preceding two seconds. The music-token condition spans the target segment, while the short latent context provides local acoustic continuity. At inference time, the preceding generated latents replace the ground-truth context, following the chunked rendering procedure in \cref{sec:music-detokenizer}.

The VAE-rate and DiT-capacity comparisons in \cref{tab:music-detokenizer-ablations} motivate the final 0.9B DiT with 50-Hz StepAudio VAE latents. After this stage, both the tokenizer and renderer are held fixed throughout LLM pretraining and post-training.

\subsection{LLM Pretraining}
\label{sec:pretraining}

The LLM in StepAudio 3 Music is initialized from a Mixture-of-Experts text language model and trained with a three-stage curriculum: large-scale pretraining, multi task mid-training, and high-quality Annealing. The curriculum first establishes alignment between lyrics and music, then teaches the model to use ABC notation as an explicit musical condition, and finally concentrates the training distribution on high-quality examples from the core creation tasks. The codec remains fixed throughout.

\paragraph{Training Corpus.}

The pre-training corpus comprises more than 100 million songs, over 5 million hours of audio, and approximately one trillion tokens. It covers instrumental and full-song generation, lyrics recognition, music understanding, cover-song generation and vocal-to-mix generation. To strengthen explicit musical control, we additionally construct a large ABC corpus containing both music-to-ABC understanding examples and ABC-to-music generation examples.

The token counts reported for the three stages denote consumed training tokens rather than mutually exclusive corpus sizes. In particular, Stage~3 replays selected high-quality material from earlier distributions. \Cref{tab:pretraining-schedule} summarizes the principal optimization settings.

\begin{table}[H]
    \centering
    \small
    \renewcommand{\arraystretch}{1.15}
    \caption{\textbf{LLM pretraining schedule for StepAudio 3 Music.}
    Token counts refer to training consumption; Stage~3 includes high-quality replay.}
    \label{tab:pretraining-schedule}
    \begin{tabular*}{\textwidth}{@{\extracolsep{\fill}}lcccl@{}}
        \toprule
        Stage & Tokens & Context & Batch & LR schedule \\
        \midrule
        Large-scale Pretraining
        & 600B & 16{,}384 & 768
        & $2.45\times10^{-5}$, constant \\
        Multi-task Mid-Pretraining
        & 500B & 32{,}768 & 512
        & $2.45\times10^{-5}\!\rightarrow\!2.0\times10^{-5}$, cosine \\
        High-quality Annealing
        & 80B & 32{,}768 & 512
        & warm-up to $2.0\times10^{-5}$, then cosine to $2.0\times10^{-6}$ \\
        \bottomrule
    \end{tabular*}
\end{table}

\paragraph{Stage~1: Large-Scale Pretraining.}

Stage~1 trains the model on approximately 600B tokens across two principal tasks. Lyrics-to-music generation accounts for roughly 500B tokens, one third of which include general text-prompt annotations. The remaining 100B tokens are allocated to music-to-lyrics recognition, providing explicit alignment between linguistic content and the music-token sequence. We use a sequence length of 16{,}384, a global batch size of 768, and a constant learning rate of $2.45\times10^{-5}$.
\paragraph{Stage~2: Multi-Task Mid-Pretraining.}

Stage~2 extends the context length and broadens the training mixture to develop generation, understanding, and reference-conditioned capabilities. The model consumes approximately 500B tokens, with lyrics-to-music generation as the primary objective and music understanding and reference-conditioned tasks providing complementary supervision. The mixture allocates approximately 32\% of tokens to standard music generation, 16\% to ABC-CoT generation, 20\% to music understanding, and 32\% to the remaining tasks.

ABC-CoT is introduced at this stage to teach the two-pass generation procedure in \cref{sec:abc-control}. Conditioned on the text prompt and lyrics, the model first produces an arrangement plan containing chords, tempo, time signature, number of bars, key, and an ABC sequence. It then generates discrete music tokens conditioned on that plan. The music-to-ABC task teaches the relationship between audio and symbolic musical structure, while ABC-to-music training teaches acoustic realization under an explicit notation condition. Together, these tasks give the LLM a structured basis for controlling melody, harmony, rhythm, and long-range form.

We use a sequence length of 32{,}768 and a global batch size of 512. The learning rate decays gradually from $2.45\times10^{-5}$ to $2.0\times10^{-5}$ following a cosine schedule.

\paragraph{Stage~3: High-Quality Annealing.}

Stage~3 concentrates training on high-quality examples from three core tasks aligned with the target application: song generation, cover-song generation, vocal-to-mix generation. The training pool contains approximately 80B tokens. Examples are selected to balance quality and diversity and are kept disjoint from the subsequent supervised fine-tuning data.

We reinitialize the optimizer and warm up the learning rate to $2.0\times10^{-5}$ over 300 steps, followed by cosine decay to $2.0\times10^{-6}$. Training uses a sequence length of 32{,}768 and a global batch size of 512. We save checkpoints after 20B, 40B, 60B, and 80B consumed tokens and evaluate them on fixed core-task validation sets and general regression suites. We select the final checkpoint based on core-task validation performance, using the general regression results to assess retention of broader capabilities.
\subsection{Post-Training}
\label{sec:posttraining}

Post-training specializes the pretrained LLM for high-quality music creation and aligns its outputs with expert listening preferences. It consists of supervised fine-tuning (SFT) followed by Direct Preference Optimization (DPO). Automatic metrics are used for data filtering and evaluation, while expert pairwise preferences supply the final alignment signal. The task mixture is centered on lyrics-to-music generation, cover-song generation and vocal-to-mix generation.

For lyrics-to-music generation, the model produces a complete song from lyrics and a text prompt. Cover-song generation preserves the melody of a reference song while changing its style; Vocal-to-mix generation takes a dry vocal track as input and produces an instrumentally arranged full mix.

\subsubsection{Supervised Fine-Tuning}
\label{sec:sft}

The SFT corpus contains approximately 20B tokens. It includes both direct music-token generation and generation preceded by explicit ABC-CoT planning. The ABC-CoT subset retains the notation plan as context for the music-token sequence, reinforcing the connection between musical conditions and their acoustic realization. \Cref{tab:sft-mixture} summarizes the currently itemized task subsets and their conditioning formats.

\begin{table}[!htbp]
    \centering
    \small
    \renewcommand{\arraystretch}{1.12}
    \caption{\textbf{Composition of the supervised fine-tuning corpus.}
    The listed subsets collectively contain approximately 20B tokens. Token counts are approximate.}
    \label{tab:sft-mixture}
    \begin{tabular*}{\textwidth}{@{\extracolsep{\fill}}lrl@{}}
        \toprule
        Subset & Tokens & Principal conditioning \\
        \midrule
        Lyrics-to-music & 6B & Lyrics and text prompt \\
        Instrumental generation & 2B & text prompt \\
        Lyrics-to-music with ABC-CoT & 4B & Lyrics, text prompt, and arrangement plan \\
        Instrumental generation with ABC-CoT & 2B & text prompt and arrangement plan\\
        Cover-song generation & 4B & Reference song \\
        Vocal-to-mix generation & 2B & Dry vocal track \\
        \bottomrule
    \end{tabular*}
\end{table}

For general music-generation examples, we group samples by genre and rank them within each group using the SongBench Musicality score \citep{wu2026songbench}. This procedure retains highly musical examples without allowing high-resource genres to dominate the selection. SFT uses a global batch size of 256 and a cosine learning-rate schedule from $2\times10^{-5}$ to $2\times10^{-6}$.

\subsubsection{Reinforcement Learning}
\label{sec:rl}

For preference alignment, we apply DPO \citep{rafailov2023direct} directly to the SFT checkpoint using an offline dataset of pairwise expert preferences. The preferred and dispreferred responses share the same generation conditions, so the optimization learns to favor outputs judged better under matched creative instructions.

The preference dataset is constructed from approximately 500 prompts selected for balanced coverage. For each prompt, we create four caption--lyrics conditioning instances and sample four audio candidates from each instance. Music annotation experts listen to and rank the candidates, considering overall musicality, melodic and harmonic coherence, condition adherence, vocal performance, and perceptual quality. Within each candidate group, we select the pair with the largest preference gap: the higher-ranked sample becomes the chosen response, and the lower-ranked sample becomes the rejected response. This procedure yields approximately 2,000 preference pairs for DPO training.

By learning under matched caption--lyrics conditions, DPO encourages the model to increase its relative preference for higher-quality candidates. Expert rankings provide a signal for perceptual distinctions that are difficult to express through existing automatic metrics. The alignment objective targets instruction following, musical coherence, and overall listening quality; its incremental effect is reserved for the controlled comparisons in \cref{tab:eval-ablations}.

\Needspace{9\baselineskip}
\section{Evaluation}
\label{sec:evaluation}

We evaluate StepAudio 3 Music through complementary objective and subjective assessments. Objective evaluation uses SongBench~\citep{wu2026songbench} for fine-grained musical quality, AudioBox-Aesthetics~\citep{tjandra2025audiobox} for perceived content and production quality, and MuQ-MuLan similarity~\citep{zhu2025muq} for caption--music alignment. Subjective evaluation uses the Vocals leaderboard of Artificial Analysis Music Arena~\citep{artificialanalysis2026music}, an independent blind-comparison benchmark. We additionally examine the contributions of acoustic rendering, preference optimization, and inference-time musical planning. These evaluations characterize generation quality; they do not directly measure adherence to individual ABC notes, chords, or bars.

\Needspace{7\baselineskip}
\subsection{Evaluation Setup}
\label{subsec:eval-setup}

The main objective comparison uses evaluator-appropriate common subsets from two tracks of our generation benchmark. AudioBox-Aesthetics and MuQ-MuLan are evaluated on the lyrics-to-song track, which contains 360 conditions. We report the fixed common subset of 339 conditions for which all systems in the original seven-system evaluation produced valid audio. SongBench is evaluated on the 316 vocal conditions in the common subset of the mixed prompt-generation track; instrumental items are omitted because SongBench's singing-related dimensions are not defined for them. Every system receives the same condition and generates one complete sample per item. Outputs missing at inference time are removed before forming the common subsets; evaluation itself produced no failed or null records.

We report comparisons with four commercial systems: Suno V5.5, Suno V5, Mureka V9, and MiniMax Music 3. The reported scores retain the original common subsets without recomputing them for the selected systems. Baseline outputs are collected through their public product interfaces within a common evaluation window; we do not assume that their service-side inference configurations can be reproduced locally. We use the default generation settings without system-specific prompt tuning.

For the main comparison, StepAudio 3 Music uses the checkpoint obtained after SFT followed by DPO, with a sampling temperature of 0.9 and top-$p$ of 0.95. The planning and editing experiments below distinguish the Direct, ABC-CoT, and ABC-CoT + LLM Editing inference modes using the DPO model. For objective evaluation, the same decoding and preprocessing pipeline is applied to every system, after which each evaluator receives the sample rate and channel layout required by its documented implementation. The arena evaluation follows the normalization and delivery procedure of Artificial Analysis rather than this internal preprocessing pipeline.

Entries in the main comparison are arithmetic means over the common subset stated for the corresponding evaluator. For MuQ-MuLan, the text input is the English prompt field, with the original prompt used only as a fallback; model-rewritten prompts are never used for scoring. Because the evaluators operate on different subsets and scales, comparisons are made between systems within a metric, not across metric families or benchmark tracks.

\subsection{Objective Evaluation}

SongBench predicts scores on a 1--10 scale along seven dimensions: Melody, Arrangement, Musicality, Vocal, Instrumental, Mixing, and Structure. The Musicality dimension captures holistic artistic impact and listening pleasure.

AudioBox-Aesthetics predicts four axes on a 1--10 scale: Content Enjoyment (CE), Content Usefulness (CU), Production Quality (PQ), and Production Complexity (PC). Higher values are better for CE, CU, and PQ. PC is descriptive: it measures the complexity of the audio scene in terms of its constituent audio components, rather than the technical quality or artistic merit of the music. Higher PC indicates a more complex scene, not necessarily better audio.

MuQ-MuLan maps music and text into a joint embedding space. We report its music--text similarity as a proxy for caption adherence; higher values indicate stronger alignment.

\begin{table}[!htbp]
    \centering
    \small
    \ADLactivate
    \renewcommand{\arraystretch}{1.14}
    \setlength{\tabcolsep}{5pt}
    \setlength{\dashlinedash}{1.4pt}
    \setlength{\dashlinegap}{2.4pt}
    \setlength{\arrayrulewidth}{0.3pt}
    \caption{\textbf{Objective evaluation on common benchmark subsets.} Caption adherence and AudioBox-Aesthetics are reported on 339 lyrics-to-song conditions; SongBench is reported on 316 vocal conditions. Values are means. Bold and underlining indicate the best and second-best scores, respectively, for metrics where higher is better. Production Complexity is descriptive and is not ranked. The dashed line separates StepAudio 3 Music from the comparison systems.}
    \label{tab:objective-main}

    \begin{tabular*}{\textwidth}{@{\extracolsep{\fill}}lc:cccc@{}}
        \toprule
        Metric
        & \shortstack{\textbf{StepAudio 3}\\\textbf{Music}}
        & \shortstack{Suno\\V5.5}
        & \shortstack{Suno\\V5}
        & \shortstack{Mureka\\V9}
        & \shortstack{MiniMax\\Music 3} \\
        \midrule

        \multicolumn{6}{l}{\textit{Caption adherence}~\citep{zhu2025muq}
        \quad (music--text similarity; $n=339$)} \\[2pt]
        MuQ-MuLan $\uparrow$
        & \textbf{0.4465} & \underline{0.4437} & 0.4287 & 0.3752 & 0.3703 \\

        \addlinespace[7pt]
        \multicolumn{6}{l}{\textit{AudioBox-Aesthetics}~\citep{tjandra2025audiobox}
        \quad (1--10; $n=339$)} \\[2pt]
        Content Enjoyment $\uparrow$
        & \textbf{7.7086} & 7.5646 & \underline{7.6823} & 7.4363 & 7.5654 \\
        Content Usefulness $\uparrow$
        & \textbf{8.0052} & \underline{7.9584} & 7.8745 & 7.6692 & 7.8677 \\
        Production Quality $\uparrow$
        & \textbf{8.3868} & \underline{8.2988} & 8.2975 & 8.0086 & 8.2976 \\
        Production Complexity
        & 6.5595 & 6.5445 & 6.5018 & 6.6607 & 6.3936 \\

        \addlinespace[7pt]
        \multicolumn{6}{l}{\textit{SongBench}~\citep{wu2026songbench}
        \quad (1--10; $n=316$)} \\[2pt]
        Melody $\uparrow$
        & 6.5394 & 6.4980 & \underline{6.6215} & \textbf{6.8121} & 6.4577 \\
        Arrangement $\uparrow$
        & \underline{6.9122} & 6.8527 & 6.8585 & \textbf{7.1490} & 6.6356 \\
        Vocal $\uparrow$
        & 7.0172 & 6.7785 & \underline{7.0412} & \textbf{7.3486} & 7.0120 \\
        Instrumental $\uparrow$
        & \underline{7.0086} & 6.8570 & 6.8934 & \textbf{7.1370} & 6.6955 \\
        Mixing $\uparrow$
        & 6.7284 & 6.8144 & \underline{6.8233} & \textbf{6.9700} & 6.5924 \\
        Structure $\uparrow$
        & \underline{6.5544} & 6.5141 & 6.5440 & \textbf{6.8821} & 6.3158 \\
        Musicality $\uparrow$
        & \underline{5.7465} & 5.5886 & 5.6966 & \textbf{5.9382} & 5.5558 \\
        \bottomrule
    \end{tabular*}
\end{table}

StepAudio 3 Music obtains the highest Content Enjoyment (7.7086), Content Usefulness (8.0052), and Production Quality (8.3868) scores under AudioBox-Aesthetics, as well as the highest MuQ-MuLan similarity (0.4465). SongBench yields a different ranking, with Mureka V9 leading its individual dimensions in this comparison. These results support strong perceived quality and caption alignment without implying uniform leadership across evaluator families. This distinction is consistent with our premise that musical organization and acoustic rendering quality are related but separate objectives.

\Cref{fig:objective-radar} summarizes the relative performance profiles of the evaluated systems across the objective metrics in \Cref{tab:objective-main}, with StepAudio 3 Music highlighted in purple.

\subsection{Subjective Evaluation}

Artificial Analysis Music Arena evaluates systems through blind pairwise comparisons of tracks generated from the same prompt. Its Quality Elo is fitted with a Bradley--Terry maximum-likelihood model and then rescaled to an Elo-like range; the leaderboard also reports 95\% confidence intervals. Before comparison, tracks are normalized toward $-16$ LUFS, capped at $-1$ dBTP, and stripped of provider metadata while preserving the original channel layout~\citep{artificialanalysis2026music}.

The arena and the internal objective evaluation differ in sample composition and task format. The internal set supplies fixed lyrics and captions, whereas the arena uses a curated prompt pool supplemented with reviewed user submissions and asks each system to generate its own lyrics. Consequently, the two evaluations provide complementary evidence but should not be treated as direct replications of one another.

\subsubsection{Artificial Analysis Benchmark Results}
\label{subsec:aa-benchmark-results}

\begin{figure}[!htbp]
    \centering
    \includegraphics[width=0.76\textwidth,height=0.46\textheight,keepaspectratio]{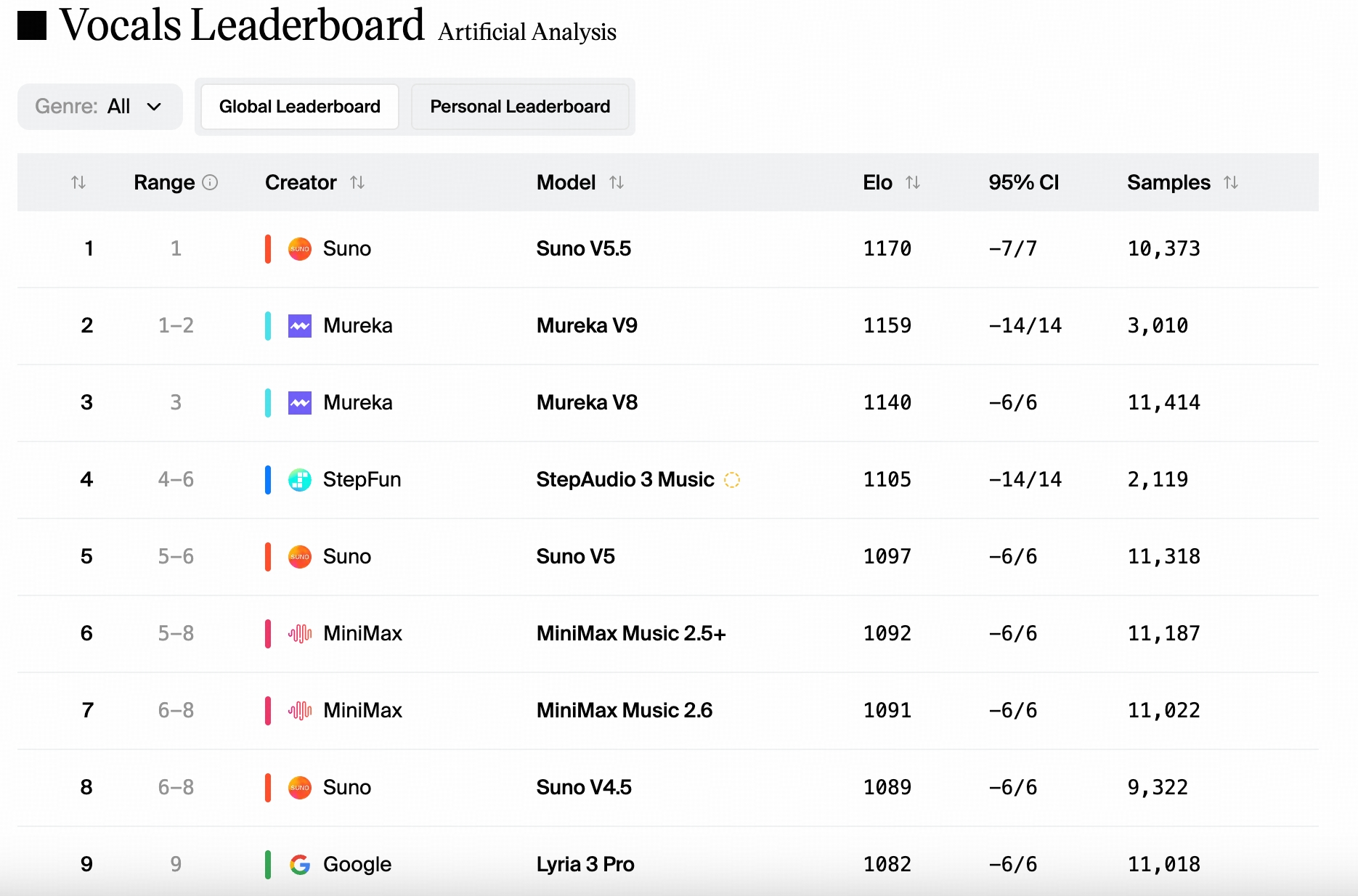}
    \caption{\textbf{Artificial Analysis Music Arena Vocals leaderboard.} Supplied preliminary benchmark results based on blind preference votes. StepAudio 3 Music obtains a Quality Elo of 1105 through direct evaluation, without any adaptation to AA prompts.}
    \label{fig:aa-benchmark}
\end{figure}

\Cref{fig:aa-benchmark} presents the supplied Artificial Analysis (AA) Music Arena Vocals leaderboard snapshot. StepAudio 3 Music achieves a Quality Elo of 1105, ranking fourth among the systems shown. These results were obtained by directly evaluating the model without any adaptation to AA prompts. The figure reproduces the supplied preliminary results.

\FloatBarrier
\Needspace{8\baselineskip}
\subsection{Ablation Studies}

We examine the system at three levels: acoustic rendering, preference optimization, and inference-time musical planning. Following the detokenizer analysis, we compare SFT with DPO to assess the post-training gains, then use the DPO model to investigate ABC-CoT planning and LLM-based plan editing.

\subsubsection{Music Detokenizer Scaling}
\label{subsec:detokenizer-ablation}

We report the reconstruction results for the VAE and DiT configurations
described in \cref{sec:music-detokenizer}. Direct VAE reconstruction and
token-conditioned rendering are evaluated separately; the DiT comparison
holds the tokenizer fixed and uses ground-truth music tokens.

\begin{table}[!htbp]
    \centering
    \footnotesize
    \renewcommand{\arraystretch}{1.12}
    \setlength{\tabcolsep}{3pt}
    \caption{\textbf{Ablations of the music detokenizer.}
    Results are reported as mean $\pm$ standard deviation over 54 audio segments. Lower MCD, MS-Mel-L1, and MS-STFT-L1 are better; higher SI-SNR, SDR, and UTMOS are better. Direct VAE reconstruction and token-conditioned detokenization are reported as separate experiment groups; UTMOS is reported only for the latter.}
    \label{tab:music-detokenizer-ablations}
    \resizebox{\textwidth}{!}{%
    \begin{tabular}{@{}lccccccc@{}}
        \toprule
        Model & Latent rate & MCD $\downarrow$ & MS-Mel-L1 $\downarrow$ & MS-STFT-L1 $\downarrow$ & SI-SNR $\uparrow$ & SDR $\uparrow$ & UTMOS $\uparrow$ \\
        \midrule
        \multicolumn{8}{l}{\emph{Direct VAE reconstruction}} \\
        StepAudio VAE (25~Hz)
        & 25~Hz & $4.79 \pm 1.31$ & $0.96 \pm 0.12$ & $0.51 \pm 0.31$ & $4.21 \pm 4.01$ & $5.54 \pm 2.96$ & -- \\
        StepAudio VAE (50~Hz)
        & 50~Hz & $3.59 \pm 1.32$ & $0.88 \pm 0.11$ & $0.42 \pm 0.26$ & $7.68 \pm 4.23$ & $8.57 \pm 3.53$ & -- \\
        \midrule
        \multicolumn{8}{l}{\emph{Token-conditioned detokenization with the StepAudio VAE (50~Hz)}} \\
        0.9B DiT
        & 50~Hz & $\mathbf{5.48 \pm 1.12}$ & $\mathbf{1.32 \pm 0.27}$ & $\mathbf{0.24 \pm 0.07}$ & $\mathbf{-19.49 \pm 9.06}$ & $-2.97 \pm 1.16$ & $\mathbf{3.45 \pm 0.60}$ \\
        4B DiT
        & 50~Hz & $6.72 \pm 0.90$ & $1.63 \pm 0.31$ & $0.27 \pm 0.11$ & $-23.27 \pm 12.23$ & $\mathbf{-2.47 \pm 1.08}$ & $3.07 \pm 0.45$ \\
        8B DiT
        & 50~Hz & $7.42 \pm 1.33$ & $1.89 \pm 0.33$ & $0.33 \pm 0.17$ & $-21.38 \pm 9.95$ & $-3.05 \pm 1.07$ & $3.20 \pm 0.42$ \\
        \bottomrule
    \end{tabular}%
    }
\end{table}

\Needspace{7\baselineskip}
\subsubsection{Preference Optimization}

We first evaluate the effect of Direct Preference Optimization (DPO) by comparing the supervised fine-tuned (SFT) model with its DPO-refined counterpart. \Cref{tab:eval-ablations} reports results across SongBench, AudioBox-Aesthetics, and MuQ-MuLan.

\begin{table}[!htbp]
    \centering
    \small
    \renewcommand{\arraystretch}{1.15}
    \setlength{\tabcolsep}{5pt}
    \caption{\textbf{Effect of DPO after supervised fine-tuning.} SB denotes SongBench, and SB-Mean is the mean over its seven dimensions. CE and PQ are the Content Enjoyment and Production Quality axes of AudioBox-Aesthetics. Higher values are better for all reported metrics.}
    \label{tab:eval-ablations}

    \begin{tabular*}{\textwidth}{@{\extracolsep{\fill}}lccccc@{}}
        \toprule
        Configuration
        & SB-Musicality $\uparrow$
        & SB-Mean $\uparrow$
        & CE $\uparrow$
        & PQ $\uparrow$
        & MuQ-MuLan $\uparrow$ \\
        \midrule
        SFT
        & 5.6395 & 6.5445 & 7.6259 & 8.2700 & 0.4157 \\
        \quad + DPO (final)
        & \textbf{5.7465} & \textbf{6.6438} & \textbf{7.7086}
        & \textbf{8.3868} & \textbf{0.4465} \\
        \bottomrule
    \end{tabular*}
\end{table}

The DPO model achieves higher scores on all five reported metrics. SongBench Musicality increases from 5.6395 to 5.7465, while the mean over its seven dimensions rises from 6.5445 to 6.6438 (+0.0993). AudioBox-Aesthetics Content Enjoyment and Production Quality improve from 7.6259 to 7.7086 and from 8.2700 to 8.3868, respectively. MuQ-MuLan similarity also increases from 0.4157 to 0.4465. Together, these results indicate improvements in musical quality, perceived audio quality, and caption alignment following preference optimization. We use the DPO model as the starting point for the planning and editing experiments below.

\Needspace{7\baselineskip}
\subsubsection{ABC-CoT Planning and LLM-Based Plan Editing}
\label{subsec:abc-cot-editing}

Building on the DPO model, we investigate whether explicit musical planning and subsequent plan refinement can further improve generation quality. We compare three inference modes: \textit{Direct}, which generates music without an explicit ABC-CoT plan; \textit{ABC-CoT}, which generates a symbolic plan before music-token generation; and \textit{ABC-CoT + LLM Editing}, which applies one round of editing by a text-based LLM to the generated plan. The editing stage refines rhythmic organization, chord progressions, and instrumentation, aiming to produce richer arrangements that better match the intended musical context. The revised ABC-CoT is then supplied to the same DPO model as a prefix for subsequent music-token generation.

\begin{table}[!htbp]
    \centering
    \small
    \renewcommand{\arraystretch}{1.15}
    \setlength{\tabcolsep}{7pt}
    \caption{\textbf{SongBench results for direct generation, ABC-CoT planning, and LLM-edited ABC-CoT planning using the DPO model.} Scores are reported on a 1--10 scale; higher is better. Mean is the arithmetic average over the seven dimensions. Bold indicates the highest value in each row.}
    \label{tab:abc-cot-editing}

    \begin{tabular*}{\textwidth}{@{\extracolsep{\fill}}lccc@{}}
        \toprule
        Dimension & Direct & ABC-CoT & ABC-CoT + LLM Editing \\
        \midrule
        Melody       & 6.5394 & 6.6054 & \textbf{6.6454} \\
        Arrangement  & 6.9122 & 6.9320 & \textbf{7.0211} \\
        Vocal        & 7.0172 & 7.0256 & \textbf{7.0282} \\
        Instrumental & 7.0086 & 7.0122 & \textbf{7.0396} \\
        Mixing       & 6.7284 & 6.7698 & \textbf{6.8483} \\
        Structure    & 6.5544 & 6.5868 & \textbf{6.6411} \\
        Musicality   & 5.7465 & 5.7676 & \textbf{5.8300} \\
        \midrule
        Mean         & 6.6438 & 6.6713 & \textbf{6.7220} \\
        \bottomrule
    \end{tabular*}
\end{table}

As shown in \cref{tab:abc-cot-editing}, ABC-CoT alone improves all seven SongBench dimensions, increasing the mean score from 6.6438 to 6.6713 (+0.0275). The gains are modest, particularly for Instrumental and Vocal. One possible explanation is that the model has not been extensively trained on explicit music-theoretic knowledge, which may limit the quality of its self-generated plans. Explicit planning provides a structured basis for generation, but its effectiveness also depends on the musical quality of the plan itself.

LLM-based editing yields a larger additional improvement, raising the mean score from 6.6713 to 6.7220 (+0.0506). The edited-plan mode improves all seven dimensions relative to ABC-CoT alone and achieves a total mean improvement of 0.0781 over Direct generation. Mean-score differences are computed before rounding. Compared with Direct, the largest absolute gains are observed in Mixing (+0.1199), Arrangement (+0.1089), and Melody (+0.1060), with further improvements in Structure (+0.0867) and Musicality (+0.0835).

These results suggest that refining the intermediate plan can help realize the benefits of explicit musical planning. ABC notation provides a readable and editable score representation through which a text-based LLM can apply music-theoretic knowledge to refine musical decisions before acoustic synthesis. By conditioning on the revised plan, the music model can translate the resulting arrangement into audio, connecting improvements in symbolic planning with improvements in generated music. Our qualitative listening observations are consistent with the measured gains, suggesting more appropriate arrangements and more enjoyable musical results. Together, these findings support interpretable symbolic plans as an interface for incorporating musical knowledge and guiding high-quality music synthesis.

\subsection{Evaluation Limitations}

SongBench is also used to filter the SFT data, so its evaluation scores are not fully independent of data selection. DPO is trained on expert pairwise preferences; the reported automatic metrics are used for filtering and evaluation. The third-party blind arena provides complementary external evidence, but it does not directly validate the ablation results because its prompts, outputs, and voters differ from those of the internal evaluation.

The main objective comparison uses one generation per condition and therefore does not measure variation across sampling seeds. Retaining the original common subsets ensures like-for-like comparisons but excludes conditions for which any system failed to produce valid audio. Baselines are evaluated through hosted services, whose complete service-side inference configurations are not available for local reproduction.

The reported subsets primarily evaluate vocal and lyrics-conditioned music generation; they do not by themselves establish comparative performance for instrumental generation, cover-song generation, or vocal-to-mix generation. These capabilities require separate task-specific evaluations. Likewise, caption similarity and overall music quality do not establish note-level or bar-level adherence to ABC notation. The planning and editing comparison in \cref{subsec:abc-cot-editing} assesses complete inference workflows; isolating the contribution of each component requires a controlled comparison. The proposed role of music-theoretic knowledge is an interpretation of the observed results, and the qualitative listening observations do not constitute a separate controlled listening study.
\FloatBarrier
\Needspace{6\baselineskip}
\section{Conclusion}
\label{sec:conclusion}

We introduced StepAudio 3 Music, a large-scale, long-form music generation model that supports explicit musical planning and open-domain text-controlled generation. The system combines a predictable music-token representation, a MoE autoregressive model, and a continuous acoustic renderer, supporting complementary forms of control through text, notation, and task-specific references.

In controlled 25-Hz comparisons, multi-codebook RVQ preserves more acoustic detail, whereas single-codebook VQ supports more stable autoregressive generation. Guided by these results, the StepAudio Music Tokenizer uses a 50-Hz, 65536-entry single codebook, with finer temporal resolution while retaining a single prediction stream. Self-supervised and multi-task training preserves structured musical content and reconstruction-relevant information. A flow-matching DiT predicts continuous StepAudio VAE latents, and our VAE decoder renders the resulting acoustic detail into 48-kHz waveform audio.

For explicit planning, the MoE model uses ABC-CoT to express an intermediate arrangement before predicting music tokens. This provides a readable, temporally organized context for melody, harmony, rhythm, and form. A progressive curriculum develops broad text- and reference-conditioned generation alongside this planning capability, followed by SFT for task execution and DPO for alignment with expert listening preferences. In objective comparisons, StepAudio 3 Music obtains the strongest AudioBox Content Enjoyment, Content Usefulness, and Production Quality scores and the highest MuQ-MuLan similarity among the evaluated systems, while remaining competitive on SongBench.

Complex arrangements, very long temporal dependencies, vocal naturalness, and high-frequency reconstruction remain areas for improvement. Future work will strengthen adherence to symbolic musical plans alongside the tokenizer and latent renderer, extending ABC-based control while improving long-form coherence and acoustic quality.

\clearpage
\small
\setlength{\bibsep}{0pt plus 0.25ex}

\bibliography{references}

\begin{thebibliography}{29}
\providecommand{\natexlab}[1]{#1}
\providecommand{\url}[1]{\texttt{#1}}
\expandafter\ifx\csname urlstyle\endcsname\relax
  \providecommand{\doi}[1]{doi: #1}\else
  \providecommand{\doi}{doi: \begingroup \urlstyle{rm}\Url}\fi

\bibitem[Evans et~al.(2024)Evans, Parker, Carr, Zukowski, Taylor, and
  Pons]{evans2024longform}
Zach Evans, Julian~D. Parker, CJ~Carr, Zack Zukowski, Josiah Taylor, and Jordi
  Pons.
\newblock Long-form music generation with latent diffusion.
\newblock \emph{arXiv preprint arXiv:2404.10301}, 2024.
\newblock URL \url{https://arxiv.org/abs/2404.10301}.

\bibitem[Ning et~al.(2025)Ning, Chen, Jiang, Hao, Ma, Wang, Yao, and
  Xie]{ning2025diffrhythm}
Ziqian Ning, Huakang Chen, Yuepeng Jiang, Chunbo Hao, Guobin Ma, Shuai Wang,
  Jixun Yao, and Lei Xie.
\newblock {DiffRhythm}: Blazingly fast and embarrassingly simple end-to-end
  full-length song generation with latent diffusion.
\newblock \emph{arXiv preprint arXiv:2503.01183}, 2025.
\newblock URL \url{https://arxiv.org/abs/2503.01183}.

\bibitem[Gong et~al.(2025)Gong, Zhao, Wang, Xu, and Guo]{gong2025acestep}
Junmin Gong, Sean Zhao, Sen Wang, Shengyuan Xu, and Joe Guo.
\newblock {ACE-Step}: A step towards music generation foundation model.
\newblock \emph{arXiv preprint arXiv:2506.00045}, 2025.
\newblock URL \url{https://arxiv.org/abs/2506.00045}.

\bibitem[Bai et~al.(2024)Bai, Chen, Chen, Chen, Deng, Dong, Hantrakul, Hao,
  Huang, Huang, et~al.]{bai2024seedmusic}
Ye~Bai, Haonan Chen, Jitong Chen, Zhuo Chen, Yi~Deng, Xiaohong Dong, Lamtharn
  Hantrakul, Weituo Hao, Qingqing Huang, Zhongyi Huang, et~al.
\newblock {Seed-Music}: A unified framework for high quality and controlled
  music generation.
\newblock \emph{arXiv preprint arXiv:2409.09214}, 2024.
\newblock URL \url{https://arxiv.org/abs/2409.09214}.

\bibitem[Zhang et~al.(2025)Zhang, Ma, Chen, Wang, Zhao, Pan, Wang, Ni, Nguyen,
  Zhou, Jiang, Tan, Gao, Du, and Ma]{zhang2025inspiremusic}
Chong Zhang, Yukun Ma, Qian Chen, Wen Wang, Shengkui Zhao, Zexu Pan, Hao Wang,
  Chongjia Ni, Trung~Hieu Nguyen, Kun Zhou, Yidi Jiang, Chaohong Tan, Zhifu
  Gao, Zhihao Du, and Bin Ma.
\newblock {InspireMusic}: Integrating super resolution and large language model
  for high-fidelity long-form music generation.
\newblock \emph{arXiv preprint arXiv:2503.00084}, 2025.
\newblock URL \url{https://arxiv.org/abs/2503.00084}.

\bibitem[Xu et~al.(2026{\natexlab{a}})Xu, Wang, Yuan, Lei, Wang, Cheng, Zhang,
  Zhang, Chen, Wang, et~al.]{xu2026qwenmusic}
Jin Xu, Kangdi Wang, Ruibin Yuan, Shun Lei, Xiong Wang, Xize Cheng, Xueyao
  Zhang, Yang Zhang, Yiheng Chen, Yongqi Wang, et~al.
\newblock {Qwen-Music} technical report.
\newblock \emph{arXiv preprint arXiv:2607.11699}, 2026{\natexlab{a}}.
\newblock URL \url{https://arxiv.org/abs/2607.11699}.

\bibitem[Gong et~al.(2026)Gong, Song, Zhao, Wang, Xu, Guo, and
  Yang]{gong2026acestep15}
Junmin Gong, Yulin Song, Wenxiao Zhao, Sen Wang, Shengyuan Xu, Jing Guo, and
  Xuerui Yang.
\newblock {ACE-Step 1.5}: Pushing the boundaries of open-source music
  generation.
\newblock \emph{arXiv preprint arXiv:2602.00744}, 2026.
\newblock URL \url{https://arxiv.org/abs/2602.00744}.

\bibitem[Dai et~al.(2026)Dai, Fan, Li, Li, Li, Ma, Ma, Ni, Shi, Tian, Wang, Wu,
  Yu, Zhang, Zhao, Zhao, and Zhu]{dai2026fullsong}
Junyu Dai, Xinyue Fan, Weiqin Li, Xiangang Li, Yunjia Li, Bin Ma, Yukun Ma,
  Chongjia Ni, Yufei Shi, Biao Tian, Haoxu Wang, Menglin Wu, Jianwei Yu,
  Huaicheng Zhang, Han Zhao, Shengkui Zhao, and Haina Zhu.
\newblock Pushing the frontier of full-song generation: Hierarchical
  autoregressive planning meets flow-matching rendering.
\newblock \emph{arXiv preprint arXiv:2607.20253}, 2026.
\newblock URL \url{https://arxiv.org/abs/2607.20253}.

\bibitem[Lam et~al.(2025)Lam, Xing, You, Wu, Yin, Jiang, Liu, Liu, Li, Lu,
  Chen, Feng, Zhao, Liu, Song, Li, and Zhou]{lam2025musiccot}
Max W.~Y. Lam, Yijin Xing, Weiya You, Jingcheng Wu, Zongyu Yin, Fuqiang Jiang,
  Hangyu Liu, Feng Liu, Xingda Li, Wei-Tsung Lu, Hanyu Chen, Tong Feng, Tianwei
  Zhao, Chien-Hung Liu, Xuchen Song, Yang Li, and Yahui Zhou.
\newblock Analyzable chain-of-musical-thought prompting for high-fidelity music
  generation.
\newblock \emph{arXiv preprint arXiv:2503.19611}, 2025.
\newblock URL \url{https://arxiv.org/abs/2503.19611}.

\bibitem[Tal et~al.(2024)Tal, Ziv, Gat, Kreuk, and Adi]{tal2024jasco}
Or~Tal, Alon Ziv, Itai Gat, Felix Kreuk, and Yossi Adi.
\newblock Joint audio and symbolic conditioning for temporally controlled
  text-to-music generation.
\newblock \emph{arXiv preprint arXiv:2406.10970}, 2024.
\newblock URL \url{https://arxiv.org/abs/2406.10970}.

\bibitem[Han et~al.(2023)Han, Dai, Hao, He, Guo, Chen, Wang, Qian, and
  Song]{han2023instructme}
Bing Han, Junyu Dai, Weituo Hao, Xinyan He, Dong Guo, Jitong Chen, Yuxuan Wang,
  Yanmin Qian, and Xuchen Song.
\newblock {InstructME}: An instruction guided music edit and remix framework
  with latent diffusion models.
\newblock \emph{arXiv preprint arXiv:2308.14360}, 2023.
\newblock URL \url{https://arxiv.org/abs/2308.14360}.

\bibitem[Yuan et~al.(2024)Yuan, Lin, Wang, Tian, Wu, Shen, Zhang, Wu, Liu,
  Zhou, et~al.]{yuan2024chatmusician}
Ruibin Yuan, Hanfeng Lin, Yi~Wang, Zeyue Tian, Shangda Wu, Tianhao Shen,
  Ge~Zhang, Yuhang Wu, Cong Liu, Ziya Zhou, et~al.
\newblock {ChatMusician}: Understanding and generating music intrinsically with
  {LLM}.
\newblock \emph{arXiv preprint arXiv:2402.16153}, 2024.
\newblock URL \url{https://arxiv.org/abs/2402.16153}.

\bibitem[Ding et~al.(2024)Ding, Liu, Dong, Zhang, Qian, He, Lin, and
  Wang]{ding2024songcomposer}
Shuangrui Ding, Zihan Liu, Xiaoyi Dong, Pan Zhang, Rui Qian, Conghui He, Dahua
  Lin, and Jiaqi Wang.
\newblock {SongComposer}: A large language model for lyric and melody
  composition in song generation.
\newblock \emph{arXiv preprint arXiv:2402.17645}, 2024.
\newblock URL \url{https://arxiv.org/abs/2402.17645v1}.

\bibitem[Li et~al.(2025)Li, Lin, Li, Huang, Wang, Wang, Zhan, and
  Wu]{li2025dualcodec}
Jiaqi Li, Xiaolong Lin, Zhekai Li, Shixi Huang, Yuancheng Wang, Chaoren Wang,
  Zhenpeng Zhan, and Zhizheng Wu.
\newblock {DualCodec}: A low-frame-rate, semantically-enhanced neural audio
  codec for speech generation.
\newblock \emph{arXiv preprint arXiv:2505.13000}, 2025.

\bibitem[{MiniMax AI}(2026)]{minimax2026music3}
{MiniMax AI}.
\newblock {MiniMax Music 3}.
\newblock \url{https://huggingface.co/MiniMaxAI/MiniMax-Music3}, 2026.
\newblock Official model card.

\bibitem[Gulati et~al.(2020)Gulati, Qin, Chiu, Parmar, Zhang, Yu, Han, Wang,
  Zhang, Wu, and Pang]{gulati2020conformer}
Anmol Gulati, James Qin, Chung-Cheng Chiu, Niki Parmar, Yu~Zhang, Jiahui Yu,
  Wei Han, Shibo Wang, Zhengdong Zhang, Yonghui Wu, and Ruoming Pang.
\newblock {Conformer}: Convolution-augmented transformer for speech
  recognition.
\newblock In \emph{Proceedings of Interspeech}, pages 5036--5040, 2020.

\bibitem[Xu et~al.(2026{\natexlab{b}})Xu, Jia, Huang, Chen, Li, Liu, Xie, Tang,
  and Hu]{xu2026fireredasr2s}
Kaituo Xu, Yan Jia, Kai Huang, Junjie Chen, Wenpeng Li, Kun Liu, Feng-Long Xie,
  Xu~Tang, and Yao Hu.
\newblock {FireRedASR2S}: A state-of-the-art industrial-grade all-in-one
  automatic speech recognition system.
\newblock \emph{arXiv preprint arXiv:2603.10420}, 2026{\natexlab{b}}.
\newblock URL \url{https://arxiv.org/pdf/2603.10420}.

\bibitem[D{\'e}fossez et~al.(2019)D{\'e}fossez, Usunier, Bottou, and
  Bach]{defossez2019music}
Alexandre D{\'e}fossez, Nicolas Usunier, L{\'e}on Bottou, and Francis Bach.
\newblock Music source separation in the waveform domain.
\newblock \emph{arXiv preprint arXiv:1911.13254}, 2019.
\newblock URL \url{https://arxiv.org/abs/1911.13254}.

\bibitem[Pratap et~al.(2023)Pratap, Tjandra, Shi, Tomasello, Babu, Kundu,
  Elkahky, Ni, Vyas, Fazel-Zarandi, Baevski, Adi, Zhang, Hsu, Conneau, and
  Auli]{pratap2023scaling}
Vineel Pratap, Andros Tjandra, Bowen Shi, Paden Tomasello, Arun Babu, Sayani
  Kundu, Ali Elkahky, Zhaoheng Ni, Apoorv Vyas, Maryam Fazel-Zarandi, Alexei
  Baevski, Yossi Adi, Xiaohui Zhang, Wei-Ning Hsu, Alexis Conneau, and Michael
  Auli.
\newblock Scaling speech technology to 1,000+ languages.
\newblock \emph{arXiv preprint arXiv:2305.13516}, 2023.
\newblock URL \url{https://arxiv.org/abs/2305.13516}.

\bibitem[Xu et~al.(2025)Xu, Xie, Tang, and Hu]{xu2025fireredasr}
Kai-Tuo Xu, Feng-Long Xie, Xu~Tang, and Yao Hu.
\newblock {FireRedASR}: Open-source industrial-grade mandarin speech
  recognition models from encoder-decoder to llm integration.
\newblock \emph{arXiv preprint arXiv:2501.14350}, 2025.
\newblock URL \url{https://arxiv.org/abs/2501.14350}.

\bibitem[Bain et~al.(2023)Bain, Huh, Han, and Zisserman]{bain2023whisperx}
Max Bain, Jaesung Huh, Tengda Han, and Andrew Zisserman.
\newblock {WhisperX}: Time-accurate speech transcription of long-form audio.
\newblock \emph{arXiv preprint arXiv:2303.00747}, 2023.
\newblock URL \url{https://arxiv.org/abs/2303.00747}.

\bibitem[Hao et~al.(2025)Hao, Yuan, Yao, Deng, Bai, Wang, Xue, and
  Xie]{hao2025songformer}
Chunbo Hao, Ruibin Yuan, Jixun Yao, Qixin Deng, Xinyi Bai, Yanbo Wang, Wei Xue,
  and Lei Xie.
\newblock {SongFormer}: Scaling music structure analysis with heterogeneous
  supervision.
\newblock \emph{arXiv preprint arXiv:2510.02797}, 2025.
\newblock URL \url{https://arxiv.org/abs/2510.02797}.

\bibitem[Chiu et~al.(2022)Chiu, Qin, Zhang, Yu, and Wu]{chiu2022bestrq}
Chung-Cheng Chiu, James Qin, Yu~Zhang, Jiahui Yu, and Yonghui Wu.
\newblock Self-supervised learning with random-projection quantizer for speech
  recognition.
\newblock In \emph{Proceedings of the 39th International Conference on Machine
  Learning}, volume 162 of \emph{Proceedings of Machine Learning Research},
  pages 3915--3924. PMLR, 2022.
\newblock URL \url{https://proceedings.mlr.press/v162/chiu22a.html}.

\bibitem[Lin et~al.(2025)Lin, Wu, Le, Wang, Chen, Dai, and
  Jiang]{lin2025duotok}
Rui Lin, Zhiyue Wu, Jiahe Le, Kangdi Wang, Weixiong Chen, Junyu Dai, and Tao
  Jiang.
\newblock {DUO-TOK}: Dual-track semantic music tokenizer for
  vocal-accompaniment generation.
\newblock \emph{arXiv preprint arXiv:2511.20224}, 2025.
\newblock URL \url{https://arxiv.org/abs/2511.20224v1}.

\bibitem[Wu et~al.(2026)Wu, Lei, Tan, Li, Wang, Zhang, Zuo, and
  Wu]{wu2026songbench}
Dapeng Wu, Shun Lei, Wei Tan, Guangzheng Li, Yunzhe Wang, Huaicheng Zhang,
  Lishi Zuo, and Zhiyong Wu.
\newblock {SongBench}: A fine-grained multi-aspect benchmark for song quality
  assessment.
\newblock \emph{arXiv preprint arXiv:2604.25937}, 2026.

\bibitem[Rafailov et~al.(2023)Rafailov, Sharma, Mitchell, Manning, Ermon, and
  Finn]{rafailov2023direct}
Rafael Rafailov, Archit Sharma, Eric Mitchell, Christopher~D Manning, Stefano
  Ermon, and Chelsea Finn.
\newblock Direct preference optimization: Your language model is secretly a
  reward model.
\newblock \emph{Advances in neural information processing systems},
  36:\penalty0 53728--53741, 2023.

\bibitem[Tjandra et~al.(2025)Tjandra, Wu, Guo, Hoffman, Ellis, Vyas, Shi, Chen,
  Le, Zacharov, Wood, Lee, and Hsu]{tjandra2025audiobox}
Andros Tjandra, Yi-Chiao Wu, Baishan Guo, John Hoffman, Brian Ellis, Apoorv
  Vyas, Bowen Shi, Sanyuan Chen, Matt Le, Nick Zacharov, Carleigh Wood, Ann
  Lee, and Wei-Ning Hsu.
\newblock {Meta Audiobox Aesthetics}: Unified automatic quality assessment for
  speech, music, and sound.
\newblock \emph{arXiv preprint arXiv:2502.05139}, 2025.

\bibitem[Zhu et~al.(2025)Zhu, Zhou, Chen, Yu, Ma, Gu, Luo, Tan, and
  Chen]{zhu2025muq}
Haina Zhu, Yizhi Zhou, Hangting Chen, Jianwei Yu, Ziyang Ma, Rongzhi Gu,
  Yi~Luo, Wei Tan, and Xie Chen.
\newblock {MuQ}: Self-supervised music representation learning with mel
  residual vector quantization.
\newblock \emph{arXiv preprint arXiv:2501.01108}, 2025.

\bibitem[{Artificial Analysis}(2026)]{artificialanalysis2026music}
{Artificial Analysis}.
\newblock Music generation benchmarking methodology and vocals leaderboard.
\newblock \url{https://artificialanalysis.ai/music/methodology}, 2026.
\newblock Vocals leaderboard available at
  \url{https://artificialanalysis.ai/music/leaderboard/vocals}; accessed 24
  August 2026.

\end{thebibliography}

\end{document}